\documentclass[12pt,a4paper,twoside,dvips]{article} 

\usepackage{My_Paper}
\title{\bf QUANTUM DYNAMICAL ENTROPIES IN DISCRETE CLASSICAL CHAOS}
\author{FABIO BENATTI\\
Dipartimento di Fisica Teorica\\
Universit\`a di Trieste\\
Strada Costiera 11, 34014 Trieste, Italy\\
and Istituto Nazionale di Fisica Nucleare, Sezione di Trieste,\\
Strada Costiera 11, 34014 Trieste, Italy\\
fabio.benatti@ts.infn.it\\
\and
VALERIO CAPPELLINI\\
Dipartimento di Fisica Teorica\\
Universit\`a di Trieste\\
Strada Costiera 11, 34014 Trieste, Italy\\
valerio.cappellini@ts.infn.it\\
\and
FEDERICO ZERTUCHE\\
Instituto de Matem\'aticas, UNAM\\
Unidad Cuernavaca, A.P. 273-3, Admon.~3\\
62251 Cuernavaca, Morelos, M\'exico.\\
zertuche@matcuer.unam.mx\\}
\begin{document}    
\maketitle
\begin{abstract}
\noindent 
We discuss certain analogies between quantization and discretization
of classical systems on manifolds. In particular, we will apply the
quantum dynamical entropy of Alicki and Fannes to numerically study
the footprints of chaos in discretized versions of hyperbolic maps on
the torus. 
\end{abstract}
\noindent Short Title: {\it Quantum Dynamical
Entropies in Discrete Classical Chaos}

\noindent Keywords: {\it Chaos, Symbolic Dynamics, 
Kolmogorov Entropy, Algorithmic Information}
\clearpage
%
\pagestyle{fancy}
\renewcommand{\headrulewidth}[0]{0pt}
\lhead[\fancyplain{}{\footnotesize \thepage \hspace{5mm}{\it
F. Benatti, V. Cappellini and F. Zertuche}}]%
      {\fancyplain{}{}}
\rhead[\fancyplain{}{}]%
      {\fancyplain{}{\footnotesize {\it Quantum Dynamical
Entropies in Discrete Classical Chaos}\hspace{5mm}\thepage}}
\chead{}\lfoot{}\cfoot{}\rfoot{}
\tableofcontents
\section{Introduction.}\label{intro}

Classical chaos is associated with motion on a compact phase--space
with high sensitivity to initial conditions: trajectories diverge
exponentially fast and nevertheless remain confined to bounded 
regions~\cite{Ref_1,Ref_2,Ref_3}.

In discrete times, such a behaviour is characterized by
a positive Lyapounov exponent $\log\lambda$, $\lambda>1$, and by
a consequent spreading of initial errors $\delta$ such that, after
$n$ time--steps,
$\delta\mapsto\delta_n\simeq\delta\,\lambda^n$.
Exponential amplification on a compact phase--space cannot grow
indefinitely, therefore the Lyapounov exponent can only be obtained as:
\[
\log\lambda \coleq \lim_{t\to\infty}\frac{1}{n}\lim_{\delta\to0}\log\left(\frac
{\delta_n}{\delta}\right)\ ,
\]
that is by first letting $\delta\to0$ and only
afterwards $n\to\infty$.

In quantum mechanics non--commutativity 
entails absence of continuous trajectories or, semi--classically,
an intrinsic coarse-graining of phase--space determined by Planck's constant
$\hbar$: this forbids $\delta$ (the minimal error possible) to go to zero.
Thus, if chaotic behaviour is identified with
$\log\lambda>0$, then it is quantally suppressed, unless,
performing the classical limit first, we let room
for $\delta\to0$~\cite{Ref_4}.

In discrete classical systems, one deals with discretized versions 
of continuous classical systems or with cellular
automata~\cite{Ref_8,Ref_9,Ref_10} with finite
number of states. In this case, roughly speaking, the minimal distance
between two 
states or configurations is strictly larger than zero; therefore, the
reason way $\log\lambda$ is trivially zero 
is very much similar to the one encountered  
in the field of quantum chaos, its origin being now not in
non--commutativity but in the lack of a continuous structure. 
Alternative methods have thus to be
developed in order to deal with the granularity of
phase--space~\cite{Ref_5,Ref_7,Ref_8,Ref_9,Ref_10}. 

An entropic approach is likely to offer a promising perspective.
For sufficiently smooth classical continuous systems, the exponential
spreading of errors is equivalent to a net entropy production,
better known as Kolmogorov dynamical, 
or \enfasi{metric}, entropy~\cite{Ref_3}.
The phase--space is partitioned
into cells by means of which any trajectory is encoded into a sequence
of symbols. 
As times goes on, the richness in different symbolic trajectories  
reflects the irregularity of the motion and is
associated with strictly positive dynamical entropy~\cite{Ref_11}.

A frequency approach to the numerical evaluation of the entropy
production has recently been applied in discretized version of various
chaotic continuous classical dynamical systems~\cite{Pigolotti}.
In this paper, we suggest a different strategy: motivated by the
similarities between quantization and discretization of continuous
classical dynamical systems, we propose to use the quantum dynamical
entropy recently introduced by Alicki and
Fannes~\cite{Ref_20,libroAF}, which we shall refer to as ALF--entropy.

The ALF--entropy is based on the algebraic properties of dynamical
systems, that is on the fact that, independently on whether they are
commutative or not, they are describable by suitable algebras of
observables, their time evolution by linear maps on these algebras and
their states by expectations over them.

As such, the ALF--entropy applies equally well to classical and
quantum systems, and reduces to the Kolmogorov entropy in the former
case. In particular, it has been showed that it allows a quite
straightforward calculation of the Lyapounov exponents of Arnold cat
maps on D--dimensional tori~\cite{Ref_30}.

In this paper we aim at showing how the ALF--entropy may be of use
in a discrete classical context, precisely when $2$--dimensional cat
maps are forced to 
live on a square lattice with spacing $\frac{1}{N}$ ($N$ integer),
with particular focus upon the emergence of the continuous behaviour
when $N\longmapsto\infty$. 

In quantum mechanics, the classical limit is achieved when $\hbar\to0$.
Analogously, in the case of discrete classical systems, by letting the
minimal distance between states go to zero, one might hope to recover a
well defined continuous dynamical system, perhaps a chaotic one.
Also, very much as in the semi--classical approximation,   
one expects the possibility of mimicking the behaviour of discrete systems 
by means of that of their continuous limits and vice versa.
However, this is possible only up to a time $\tau_B$,
called \enfasi{breaking--time}~\cite{Ref_4}; it can be heuristically
estimated as the time when
the minimal error permitted, $\delta$, becomes of the order of the 
phase--space bound $\Delta$. Therefore, when, in the continuum, a 
Lyapounov exponent $\log\lambda>0$ is present, the breaking--time scales as  
$\displaystyle
\tau_B=\frac{1}{\log\lambda}\log\frac{\Delta}{\delta}$ .

In the following, we shall consider discrete dynamical systems
obtained by discretizing
a subclass
of the Unitary Modular Group of ($2$--dimensional) Toral
Automorphisms~\cite{Ref_3}  
containing the well known Arnold Cat Maps. We shall provide:
\begin{itemize}
\item the algebraic setting for the continuous limit $N\longmapsto\infty$;
\item the technical framework to construct the ALF--entropy 
      and numerical shortcuts to compute it;
\end{itemize}
and study:
\begin{itemize}
\item the behaviour of the entropy production and how the
      breaking--time $\tau_B$ is reached in hyperbolic 
      systems;
\item the differences in behavior between hyperbolic and elliptic
      systems;
\item the distribution of eigenvalues of the multitime correlation matrix
      used in computing the ALF--entropy;
\item the behaviour of the entropy production in the case of
Sawtooth Maps~\cite{Ref_24,Ref_25,Ref_26} which are discontinuous on
the $2$--dimensional torus. 
\end{itemize}  

\section{Automorphisms on the Torus}\label{AATT}

Usually, continuous classical motion is described by means of a
measure space ${\cal X}$, the phase--space, endowed with the Borel
$\sigma$--algebra and a normalized 
measure $\mu$, $\mu({\cal X})=1$.
The ``volumes'' 
$\displaystyle
\mu(E)=\int_E{\rm d}\mu(x)$ of measurable subsets 
$E\subseteq{\cal X}$ represent the probabilities that a
phase--point $x\in{\cal X}$ belong to them.
By specifying the statistical properties of the system, the measure $\mu$ 
defines a ``state'' of it.

In such a scheme, a reversible discrete time dynamics amounts to an
invertible measurable 
map $T:{\cal X}\mapsto{\cal X}$ such that $\mu\circ T=\mu$ and to its
iterates $\{T^j\}_{j\in\IZ}$.
Phase--trajectories passing through $x\in{\cal X}$ at time $0$
are then sequences $\{T^j x\}_{j\in\IZ}$~\cite{Ref_3}.

Classical dynamical systems are thus conveniently described by
triplets $({\cal X},\mu,T)$; in the following, we shall concentrate
on triplets $({\cal X},\mu,T_\alpha)$, where 
\begin{subequations}
\label{AoDC_1}
\begin{align}
{\cal X}&={\IT}^2={\IR}^2/{\IZ}^2=\left\{\bs{x}=(x_1,x_2)\ \pmod{1} \right\}
\label{AoDC_1a}\\
T_\alpha 
\begin{pmatrix}
x_1\\
x_2
\end{pmatrix}& =
\begin{pmatrix}
1+\alpha & 1\\
\alpha & 1
\end{pmatrix}
\begin{pmatrix}
x_1\\
x_2
\end{pmatrix}
\ \pmod{1}\ ,\quad
\alpha\in\IZ
\label{AoDC_1b}\\
\ud\mu(\bs{x})&=\ud x_1\; \ud x_2 \ \cdot
\label{AoDC_1c}
\end{align} 
\end{subequations}
\ \\[-8.5ex]
\begin{quote}
\begin{NNS}{}\ \\[-5ex]\label{Rem_21}
\begin{Ventry}{\mdseries iii.}
\item[\mdseries i.] Since $\det\pt{T_\alpha}=1$, the Lebesgue measure
defined in~\eqref{AoDC_1c} is \enfasi{invariant} for all $\alpha\in\IZ$;
\item[\mdseries ii.] The eigenvalues of 
	$\pt{\begin{smallmatrix} 1+ \alpha  & 1\\ \alpha & 1 
	\end{smallmatrix}}$ are  
 	$(\alpha+2\pm\sqrt{(\alpha+2)^2-4})/2$.
	They are conjugate complex
	numbers if
	$\alpha\in\pq{-4,0}$, while one eigenvalue $\lambda$ is
	greater than $1$ if $\alpha\not\in\pq{-4,0}$.
	In this case,
	distances are stretched along the direction of the
	eigenvector $|\bs{e}_+\rangle$,
	$S_\alpha|\bs{e}_+\rangle=\lambda|\bs{e}_+\rangle$,
	contracted along 
	that of $|\bs{e}_-\rangle$, $S_\alpha|\bs{e}_-\rangle=
	\lambda^{-1}|\bs{e}_-\rangle$.\\
	For such $\alpha$'s all periodic points are
	hyperbolic~\cite{Ref_26}.	
\item[\mdseries iii.] 
	$T_1=\pt{\begin{smallmatrix} 2 & 1\\ 1 & 1
	\end{smallmatrix}}$ is the Arnold Cat Map~\cite{Ref_3}.
	Then, $\displaystyle T_1 \in 
	{\left\{T_{\alpha}\right\}}_{\alpha\in\IZ} \subset
	{\text{SL}}_2\left(\IT^2\right)\subset
	{\text{GL}}_2\left(\IT^2\right)\subset
	{\text{ML}}_2\left(\IT^2\right)$ where
	${\text{ML}}_2\left(\IT^2\right)$ is the subset of $2\times2$
	matrices with integer entries, 
	${\text{GL}}_2\left(\IT^2\right)$ the subset of
	invertible matrices and ${\text{SL}}_2\left(\IT^2\right)$ the subset
	of matrices with determinant one.
\item[\mdseries iv.] The dynamics generated by
	$T_{\alpha}\in{\text{SL}}_2\left(\IT^2\right)$
	is called \enfasi{Unitary Modular Group}~\cite{Ref_3} (UMG for short).
\end{Ventry}
\end{NNS}
\end{quote}
\noindent
For future comparison with quantum dynamical systems, we adopt
an algebraic point of view and argue in terms of classical 
observables, precisely in terms of complex continuous functions $f$ on
${\cal X}={\IT}^2$. 
\begin{itemize}
	\item These functions form a C* algebra ${\cal A}_{{\cal X}}=\Cspace{0}{\cal X}$ with respect to the
topology given by the \enfasi{uniform norm} $\displaystyle
\norm{f}{0}=\sup_{\bs{x}\in{\cal X}}\Big|f\pt{\bs{x}}\Big|$;
	\item the Lebesgue measure $\mu$ defines a state $\omega_\mu$ on ${\cal
A}_{\cal X}$ which evaluates mean values of observables via
integration: 
\begin{equation}
f\mapsto\omega_\mu(f)\coleq\int_{{\cal X}} \ud x\ f(x)\ ;
\label{AoDC_2}
\end{equation}
	\item the discrete--time dynamics $T_\alpha: {\cal X} \mapsto {\cal X}$
generates the discrete group of automorphisms 
$\Theta_{\alpha}^{j}: {\cal A}_{{\cal X}}\mapsto {\cal A}_{{\cal X}}$,
given by 
\begin{equation}
\Theta_{\alpha}^j\pt{f}\pt{\bs{x}}=f(S_{\alpha}^j \pt{\bs{x}})\ , \quad j\in \IZ\ , 
\label{AoDC_3}
\end{equation}
that preserve the state,
$\omega_\mu\circ\Theta_{\alpha}^j=\omega_\mu$. 
\end{itemize}
\ \\[-7.5ex]
\begin{quote}
\begin{DDD}{}\ \\[-5ex]
\begin{Ventry}{}\label{RoeiW_51c}
	\item[] The dynamical systems $\pt{{\cal X},\mu,T_\alpha}$ will
	be identified by the algebraic triplets
	$\pt{{\cal A}_{\cal X}, \omega_\mu, \Theta_\alpha}$. 
\end{Ventry}
\end{DDD}
\end{quote}
\noindent
\subsection{``Weyl'' discretization}\label{roL}
In the following we shall proceed to a discretization of the systems
introduced in the previous Section and to the study of
how chaos emerges 
when the continuous limit is
being reached.

Roughly speaking, given an integer $N$, we shall force the continuous
classical systems 
$\pt{{\cal A}_{\cal X}, \omega_\mu, \Theta_\alpha}$ 
to live on a lattice $L_N\subset 
{\IT}^2$ given by:
\begin{equation}
L_N \coleq \pg{\frac{\bs{p}}{N} \ \Big|\  \bs{p}\in {\pt{\IZ / N \IZ}}^2}\ ,
\label{llnn}
\end{equation}
where $\pt{\IZ / N \IZ}$ denotes the residual
class$\pmod{N}$.

A good indicator of chaos in continuous dynamical systems is the
metric entropy of Kolmogorov~\cite{Ref_3}~(see Section~\ref{KSE} below).
We can compare discretization of classical continuous
systems with quantization; 
in this way, we can profitably
use a quantum extension of the metric entropy  which will be
presented in Section~\ref{AFE}. 
To this aim, we define a
discretization procedure resembling Weyl quantization~\cite{Ref_27,Ref_28}; 
in practice, we will construct a *morphism ${\cal J}_{N , \infty}$
from 
${\cal A}_{\cal X} = \Cspace{0}{\cal X}$ into the abelian algebra
$D_{N^2}\pt{\IC}$ of $N^2 \times 
N^2$ matrices which are diagonal with respect to a chosen
orthonormal basis ${\pg{\ket{\bs{\ell}}}}_{\bs{\ell}\in 
{\pt{\IZ / N \IZ}}^2}$. The basis vectors
will be labeled by the points of a square grid of lattice
spacing 
$\frac{1}{N}$ with $0 \leqslant \ell_i \leqslant \pt{N-1}$ ($N$
identified with $0$) superimposed onto ${\cal X} = {\IT}^2$.

In order to define ${\cal J}_{N , \infty}$, we use Fourier analysis
and restrict ourselves to the
*subalgebra ${\cal W}_{\text{exp}}\in{\cal A}_{\cal X}$ generated
by the exponential functions
\begin{equation}
W(\bs{n})(\bs{x})=\exp(2\pi i\:\bs{n}\cdot\bs{x})\ , \label{equ_22}
\end{equation}
where $\bs{n}=(n_1,n_2)\in\IZ^2$ and
\mbox{$\bs{n}\cdot\bs{x}=n_1\,x_1+n_2\,x_2$}.
The generic element of ${\cal W}_{\text{exp}}$ is: 
\begin{equation}
f(\bs{x}) = \sum_{\bs{n} \in {\IZ}^2} \hat{f}_{\bs{n}}
W(\bs{n})(\bs{x})\label{RoeiW_1}
\end{equation} with finitely many coefficients $\displaystyle
\hat{f}_{\bs{n}} = \int\!\!\!\!\!\int_{\cal X} \ud
\bs{x} \; f (\bs{x}) \; e^{- 2\pi i \bs{n}\bs{x}}$ different
from zero.\\
On ${\cal W}_{\text{exp}}$, formula~\eqref{AoDC_2} defines a state such that
\begin{equation}
	\omega_\mu\left(W(\bs{n})\right)=\delta_{\bs{n},\bs{0}} 
\label{fabio1}\cdot    
\end{equation}
Further, since $\bs{n}\cdot\pt{T_\alpha^{\phantom{tr}}\bs{x}}=
\pt{T_\alpha^{tr}\bs{n}}\cdot\bs{x}$, the automorphisms~\eqref{AoDC_3}
map exponentials into themselves:
\begin{equation} 
\Theta_{\alpha}\left(W(\bs{n})\right) = W(T^{\text{tr}}_{\alpha}\cdot\bs{n})\ ,
\qquad T_{\alpha}^{\text{tr}}=\begin{pmatrix}1+\alpha & \alpha\\ 1&
1\end{pmatrix}
\label{fabio2}
\end{equation}\\[-7.5ex]
\begin{quote}
\begin{NNN}{}\ \\[-0.5ex]
\label{ssssmmmm}
The latter property  no longer holds 
when $\alpha\not\in\IZ$ as will be the case in
Section~\ref{Sawtooth} where we deal with Sawtooth
Maps~\cite{Ref_24,Ref_25,Ref_26}.
\end{NNN}
\end{quote}
\noindent
Following Weyl quantization, we get elements of
$D_{N^2}$ out of elements  of ${\cal W}_{\text{exp}}$ by replacing,
in~\eqref{RoeiW_1}, exponentials with 
diagonal matrices:
\begin{equation}
W(\bs{n}) \longmapsto \widetilde{W}(\bs{n}) \coleq \sum_{\bs{\ell} \in {(\ZNZ{N})^2}} 
e^\frac{\:2\pi i \bs{n}\bs{\ell}}{N}
\ket{\bs{\ell}}\bra{\bs{\ell}} \ ,\qquad
\bs{\ell} =\pt{\ell_1 , \ell_2}\cdot\label{RoeiW_4}
\end{equation}
\\[-2ex]
\begin{quote}
\begin{DDD}{}\ \\[-5ex]
\begin{Ventry}{}\label{RoeiW_51}
	\item[] We will denote by ${\cal J}_{N ,
	\infty}^{\cal W}$, the *morphism from the 
	\mbox{*algebra} ${\cal W}_{\text{exp}}$ into the diagonal matrix algebra
	$D_{N^2}\pt{\IC}$, given by:
\begin{align}
{\cal W}_{\text{exp}}\ni f \longmapsto {\cal J}_{N , \infty}^{\cal W}(f) &
\coleq \sum_{\bs{n} \in {\IZ}^2} \hat{f}_{\bs{n}}
\;\widetilde{W}(\bs{n})\notag \\
& = \sum_{\bs{\ell} \in {(\ZNZ{N})^2}} 
f\pt{\frac{\bs{\ell}}{N}}
\ket{\bs{\ell}}\bra{\bs{\ell}}\cdot\label{RoeiW_5}
\end{align} 
\end{Ventry}
\end{DDD}
\end{quote}
\noindent
\\[-7.5ex]
\begin{quote}
\begin{NNS}{}\ \\[-5ex]
\begin{Ventry}{iii)}
	\item[\mdseries i)] The completion of the subalgebra ${\cal
	W}_{\infty}$ with respect to the uniform norm
	$\displaystyle \norm{f}{0}=\sup_{\bs{x}\in{\cal
	X}}\Big|f\pt{\bs{x}}\Big|$ is the C* algebra ${\cal A}_{\cal  
	X} = \Cspace{0}{\cal X}$~\cite{Ref_28bis}.
	\item[\mdseries ii)] 
	The *morphism ${\cal J}_{N ,
	\infty}^{\cal W}:{\cal W}_{\text{exp}}\mapsto D_{N^2}\pt{\IC}$ is
	bounded by \mbox{$\norm{{\cal J}_{N ,
	\infty}^{\cal W}}{} = 1$.} Using the Bounded Limit
	Theorem~\cite{Ref_28bis}, ${\cal J}_{N
	,\infty}^{\cal W}$ can be uniquely extended to a bounded linear
	transformation (with the same bound) ${\cal J}_{N
	,\infty}: {\cal A}_{\cal X} \longmapsto D_{N^2}\pt{\IC}$.
	\item[\mdseries iii)] ${\cal J}_{N,\infty}\pt{{\cal A}_{\cal X}}
	= D_{N^2}\pt{\IC}$.
\end{Ventry}
\end{NNS}
\end{quote}
\noindent
We go back from $D_{N^2}\pt{\IC}$ to a *algebra of functions on ${\cal
X}$ by defining a *morphism
${\cal J}_{\infty , N}$ that ``inverts'' ${\cal J}_{N ,
\infty}$ in the $N \to \infty$ limit. In 
the Weyl quantization the ``inverting'' *morphism is constructed by
means of
coherent states $\ket{\beta\pt{\bs{x}}}$, $\bs{x}\in {\cal X}$,  with
good localization properties in ${\cal X}$.\\[-2ex]
\begin{quote}
\begin{DDD}{}\ \\[-5ex]
\begin{Ventry}{}\label{RoeiW_51b}
	\item[] We will denote by ${\cal J}_{\infty
	,N}:D_{N^2}\pt{\IC}\longmapsto {\cal A}_{\cal X} $ the *morphism 
	defined by:
\begin{equation}
D_{N^2}\pt{\IC}\ni M \longmapsto{\cal J}_{\infty , N}(M)\pt{\bs{x}}
\coleq \bra{\beta\pt{\bs{x}}} M \ket{\beta\pt{\bs{x}}}\ ,
\label{RoeiW_7}
\end{equation}
where $\ket{\beta\pt{\bs{x}}}$ are coherent vectors in ${\cal
H}_{N^2}= {\IC}^{N^2}$
\end{Ventry}
\end{DDD}
\end{quote}
\noindent
We now construct a suitable family of $\ket{\beta\pt{\bs{x}}}$:
we shall denote by $\floor{\cdot}$ and $\bk{\cdot}$ the integer and
fractional part of a real number so that 
we can express each ${\IT}^2$
as $\bs{x} =\pt{\frac{\floor{N
x_1}}{N},\frac{\floor{N x_2}}{N}} + 
\pt{\frac{\bk{N x_1}}{N},\frac{\bk{N x_2}}{N}}$. Then we associate
$\bs{x}\in{\IT}^2$ with vectors of ${\cal H}_{N^2}$ as follows:  
\begin{multline}
\label{RoeiW_6}
\bs{x} \mapsto \ket{\beta\pt{\bs{x}}} = 
\lambda_{11}\pt{\bs{x}}\ket{\floor{N x_1},\floor{N
x_2}} +\\ 
+ \lambda_{12}\pt{\bs{x}}\ket{\floor{N x_1},\floor{N
x_2}+1} +
\lambda_{21}\pt{\bs{x}}\ket{\floor{N x_1}+1,\floor{N
x_2}} +\\ 
+ \lambda_{22}\pt{\bs{x}}\ket{\floor{N x_1}+1,\floor{N
x_2}+1} \cdot
\end{multline}
We choose the coefficients $\lambda_{ij}$ so that
$\norm{\beta\pt{\bs{x}}}{} = 1$ and that the map~\eqref{RoeiW_6} be
invertible: 
\begin{equation}
\begin{cases}
\lambda_{11}\pt{\bs{x}}= \cos\pt{\frac{\pi}{2}\bk{N x_1}} 
\cos\pt{\frac{\pi}{2}\bk{N x_2}}\\
\lambda_{12}\pt{\bs{x}}= \cos\pt{\frac{\pi}{2}\bk{N x_1}} 
\sin\pt{\frac{\pi}{2}\bk{N x_2}}\\
\lambda_{21}\pt{\bs{x}}= \sin\pt{\frac{\pi}{2}\bk{N x_1}} 
\cos\pt{\frac{\pi}{2}\bk{N x_2}}\\
\lambda_{22}\pt{\bs{x}}= \sin\pt{\frac{\pi}{2}\bk{N x_1}} 
\sin\pt{\frac{\pi}{2}\bk{N x_2}}
\end{cases}
\label{RoeiW_65}
\end{equation}
Therefore, from definitions~\ref{RoeiW_51} and \ref{RoeiW_51b} it
follows that, when mapping ${\cal A}_{\cal X}$ onto
$D_{N^2}\pt{\IC}$ and the latter back into ${\cal A}_{\cal X}$, we get:
\begin{multline}
\widetilde{f}_N\pt{\bs{x}}\coleq
\pt{{\cal J}_{\infty , N}\circ {\cal J}_{N , \infty}}
\pt{f}\pt{\bs{x}} =
\sum_{\bs{\ell} \in {(\ZNZ{N})^2}} 
f\pt{\frac{\bs{\ell}}{N}}
{\abs{\bk{ \beta\pt{\bs{x}}\big| \bs{\ell}}}}^2 =\\
= \frac{1}{4}\sum_{\pt{\mu,\nu,\rho,\sigma}\in{\pg{0,1}}^4} 
\cos\pt{\pi \mu \bk{N x_1}} \cos\pt{\pi \nu \bk{N x_2}}
{\pt{-1}}^{\mu \rho + \nu \sigma} \times\\
\times f\pt{\frac{\floor{N x_1}+\rho}{N},\frac{\floor{N x_2}+\sigma}{N}}
\label{RoeiW_8}
\end{multline}
\begin{quote}
\begin{NNS}{}\ \\[-5ex]
\label{rem_24}
\begin{Ventry}{ii)}
	\item[\mdseries i)] From~\eqref{RoeiW_8},
	$f = \widetilde{f}_N$ on
	the lattice points. Moreover, although the first derivative
	of~\eqref{RoeiW_8} is not defined on the latter, its
	limit exists there and it is zero; thus, we can extend by continuity 
	$\widetilde{f}_N$ to a function in $\Cspace{1}{\IT^2}$ that we
	will denote again as $\widetilde{f}_N$.
	\item[\mdseries ii)] We note that $\mathrm{Ran}\pt{{\cal
	J}_{\infty , N}}$ is a subalgebra 
 	strictly contained in ${\cal A}_{\cal X}$; this is not
	surprising and comes as consequence of Weyl quantization,
	where this phenomenon is quite
	typical~\cite{Ref_27,Ref_28}.
\end{Ventry}
\end{NNS}
\end{quote}
\noindent
We show below that \mbox{${\cal J}_{\infty , N} \circ {\cal J}_{N ,
\infty}$} approaches ${\Id}_{{\cal A}_{\cal
X}}$ (the identity function in ${\cal A}_{\cal
X}$) when $N \to \infty$. 
Indeed, a request upon any sensible quantization procedure is
to recover 
the classical description in the limit $\hbar \to 0$; in a
similar way, our discretization should recover the continuous
system in the $\frac{1}{N}\to 0$ limit. 
\begin{TT}
Given $f \in {\cal A}_{\cal X}=\Cspace{0}{\IT^2},$
\mbox{$\displaystyle
\  \lim_{N\to\infty}{\bigg\|\pt{{\cal J}_{\infty , N} \circ {\cal
J}_{N , \infty}-{\Id}_{{\cal A}_{\cal X}}}\pt{f}\bigg\|} = 0
\; \cdot
$}
\end{TT}
\noindent
{\bf Proof:}\quad Since ${\cal X}
= \IT^2$ is compact, $f$ is uniformly continuous on it.
Further, denoting $\displaystyle \bs{x}=\frac{\floor{N\bs{x}}}{N} +
\frac{\bk{N\bs{x}}}{N}$, $0\leq \bk{N\bs{x}} <1$ implies 
$\displaystyle \norm{\bs{x} -
\frac{\floor{N\bs{x}}}{N}}{}\xrightarrow[\ N\ ]{} 0$; therefore, for all
$\varepsilon>0$ there exists $N_{f,\varepsilon}$ such that
\begin{equation*}
N>\bar{N}_{f,\varepsilon} \Longrightarrow 
\:\abs{f\pt{\bs{x}} - f\pt{\frac{\floor{N\bs{x}}}{N}}\:} <
\frac{\varepsilon}{2}\ ,\notag
\end{equation*}
uniformly in $\bs{x}$. Moreover, according to Remark~\ref{rem_24}~(i.),
$\widetilde{f}_N\in\Cspace{0}{\IT^2}$, thus the previous inequality holds
for $\widetilde{f}_N$, too. Since $\widetilde{f}_N=f$ on $\displaystyle
\frac{\floor{N\bs{x}}}{N}$, it follows that, for sufficiently large
$N$, 
\begin{equation*}
 \abs{f\pt{\bs{x}}-\widetilde{f}_N \pt{\bs{x}}}\leq
\abs{f\pt{\bs{x}} - f\pt{\frac{\floor{N\bs{x}}}{N}}} + 
\abs{\widetilde{f}_{N}\pt{\frac{\floor{N\bs{x}}}{N}} -
\widetilde{f}_{N}\pt{\bs{x}}} 
\leq \varepsilon\ ,\notag
\end{equation*}
uniformly in $\bs{x}$.\hfill$\qed$
\section{Kolmogorov metric entropy}\label{KSE}
For continuous classical systems $\tripC$ such as
those introduced in 
Section~\ref{AATT}, the construction of the dynamical entropy of
Kolmogorov is based  
on 
subdividing $\cal X$ into measurable disjoint subsets
${\left\{E_\ell\right\}}_{\ell=1,2,\cdots, D}$ such that $\bigcup_\ell
E_\ell={\cal X}$ which form finite partitions (coarse grainings) 
${\cal E}$.

Under the dynamical maps $T_{\alpha}$ in~\eqref{AoDC_1b},
any given ${\cal E}$ evolves into $T_{\alpha}^{-j}({\cal E})$ with atoms
$\displaystyle
T_{\alpha}^{-j}(E_\ell)=\{x\in{\cal X}: T_{\alpha}^jx\in E_\ell\}$;
one can then form finer partitions
$\displaystyle{\cal E}_{[0,n-1]}$
whose atoms
$\displaystyle
E_{i_0\,i_1\cdots i_{n-1}}\coleq E_{i_0}\bigcap T_{\alpha}^{-1}(E_{i_1})\cdots\bigcap 
T_{\alpha}^{-n+1}(E_{i_{n-1}})$
have volumes
\begin{equation}
\mu_{i_0\,i_1\cdots i_{n-1}}\coleq\mu\left(E_{i_0}\cap T_{\alpha}^{-1}(E_{i_1})\cdots
\cap T_{\alpha}^{-n+1}(E_{i_{n-1}})\right)\ . \label{KSE_1}
\end{equation}\\[-7.5ex]
\begin{quote}
\begin{DDD}{}\label{stringhe}\ \\[-0.5ex]
We shall set $\bs{i}=\pg{i_0\,i_1\cdots i_{n-1}}$ and
denote by $\Omega_D^n$ the set of $D^n$ n\_tuples with $i_j$ taking
values in $\pg{1, 2, \cdots, D}$.
\end{DDD}
\end{quote}
\noindent
The atoms of the partitions ${\cal E}_{[0,n-1]}$ describe segments 
of trajectories 
up to time $n$ encoded by the atoms of ${\cal E}$ that are traversed at 
successive times.
The richness in diverse trajectories, that is the degree of irregularity 
of the 
motion (as seen with the accuracy of the given coarse-graining), 
can be measured by the Shannon entropy~\cite{Ref_11} 
\begin{equation}
S_\mu({\cal E}_{[0,n-1]})\coleq-\sum_{\bs{i}\in\Omega_D^n}\mu_{\bs{i}}
\log\mu_{\bs{i}}\ .
\label{KSE_2}
\end{equation}
On the long run, ${\cal E}$ attributes to the dynamics an entropy per 
unit time--step
\begin{equation}
h_\mu(T_{\alpha},{\cal E})\coleq\lim_{n\to\infty}\frac{1}{n}S_\mu({\cal E}_{[0,n-1]})\ .
\label{KSE_3}
\end{equation}
This limit is well defined~\cite{Ref_3} and the Kolmogorov
entropy $h_\mu(T_{\alpha})$ of $\pt{{\cal A}_{\cal
X},\omega_\mu,\Theta_\alpha}$
is defined
as the supremum over all finite measurable
partitions~\cite{Ref_3,Ref_11}:
\begin{equation}
h_\mu(T_{\alpha})\coleq\sup_{{\cal E}}h_\mu(T_{\alpha},{\cal E})\ \cdot
\label{KSE_4}
\end{equation}
\subsection{Symbolic Models as Classical Spin Chains}
Finite partitions ${\cal E}$ of ${\cal X}$ provide symbolic models 
for the dynamical systems $\tripC$ of Section~\ref{AATT},
whereby the trajectories  
${\left\{T_{\alpha}^jx\right\}}_{j\in\IZ}$ are encoded into sequences
${\left\{i_j\right\}}_{j\in\IZ}$ 
of indices relative to the atoms $E_{i_j}$ visited at successive times $j$;
the dynamics corresponds to the right--shift along the symbolic sequences. 
The encoding can be modelled as the shift along
a classical spin chain 
endowed with a shift--invariant state~\cite{Ref_20}.
This will help to understand the quantum dynamical entropy which will be
introduced in the next Section.

Let $D$ be the number of atoms of a partition ${\cal E}$ of ${\cal
X}$, we shall denote by $\bs{A}_D$ the  
diagonal $D\times D$ matrix algebra generated by the characteristic
functions $e_{E_{\ell}}$ of the atoms $E_{\ell}$ and by
$\bs{A}_D^{[0,n-1]}$ the $n$-fold 
tensor product of $n$ copies of $(\bs{A}_D)$, that is the $D^n\times D^n$
diagonal matrix algebra
$\bs{A}_D^{[0,n-1]}\coleq(\bs{A}_D)_0\otimes(\bs{A}_D)_1\cdots
\otimes(\bs{A}_D)_{n-1}$.
Its typical elements are of the form $a_0\otimes a_1\cdots\otimes a_{n-1}$
each $a_j$ being a diagonal $D\times D$ matrix. 
Every $\bs{A}_D^{[p,q]}\coleq\otimes_{j=p}^q(\bs{A}_D)_j$ can be embedded into
the infinite tensor product 
$\displaystyle\bs{A}_D^\infty\coleq\otimes_{k=0}^\infty(\bs{A}_D)_k$ as
\begin{equation}
(\Id)_0\otimes\cdots\otimes(\Id)_{p-1}\otimes(\bs{A}_D)_p\otimes\cdots
\otimes(\bs{A}_D)_q\otimes(\Id)_{q+1}\otimes(\Id)_{q+2}
\otimes\cdots
\label{SMaCSC_1}
\end{equation}
The algebra $\bs{A}_D^\infty$ is a classical spin chain with a
classical $D$--spin at each site.

By means of the discrete probability measure
$\{\mu_{\bs{i}}\}_{\bs{i}\in\Omega_D^n}$, one can 
define a compatible family of states on the ``local'' algebras 
$\bs{A}_D^{[0,n-1]}$:
\begin{equation}
\rho_{{\cal E}}^{[0,n-1]}\left(a_0\otimes\cdots\otimes a_{n-1}\right)
=\sum_{\bs{i}\in\Omega_D^n}\mu_{\bs{i}}\,
(a_0)_{i_0i_0}\cdots
(a_{n-1})_{i_{n-1}i_{n-1}}\ .
\label{SMaCSC_2}
\end{equation}
\noindent Indeed, let $\rho\rstr\bs{N}$ denote the restriction to a subalgebra $\bs{N}
\subseteq\bs{M}$ of a state $\rho$ on a larger
algebra $\bs{M}$.
Since
$\sum_{i_{n-1}}\mu_{i_0i_1\cdots i_{n-1}}=\mu_{i_0i_1\cdots  i_{n-2}}$,
when $n$ varies the local states $\rho_{{\cal E}}^{[0,n-1]}$ are such that
$\displaystyle
\rho_{{\cal E}}^{[0,n-1]}\rstr\bs{A}^{[0,n-2]}_D=\rho_{{\cal
E}}^{[0,n-2]}$
and define a
``global'' state $\rho_{{\cal E}}$ on
$\displaystyle \bs{A}_D^\infty$
such that $\rho_{{\cal E}}\rstr\bs{A}_D^{[0,n-1]}=\rho_{{\cal
E}}^{[0,n-1]}$.

\noindent From the $T_{\alpha}$-invariance of $\mu$ it follows that,
under the right--shift
$\displaystyle
\sigma:\bs{A}_D^\infty\mapsto\bs{A}_D^\infty$,
\begin{equation}
\sigma\pt{\bs{A}_D^{[p,q]}}=\bs{A}_D^{[p+1,q+1]}\ ,
\label{SMaCSC_3}
\end{equation}
the state $\rho_{{\cal E}}$ of the classical spin chain is translation
invariant:
\begin{equation}
\begin{split}
\rho_{{\cal E}}\circ\sigma\left(a_0\otimes\cdots\otimes a_{n-1}\right)
&=\rho_{{\cal E}}\left((\Id)_0\otimes(a_0)_1\otimes\cdots\otimes (a_{n-1})_n
\right)\\
&=\rho_{{\cal E}}\left(a_0\otimes\cdots\otimes a_{n-1}\right)\ \cdot
\end{split}
\label{SMaCSC_4}
\end{equation}
Finally, denoting by $|j\rangle$ the basis vectors of the representation 
where the matrices $a\in\bs{A}_D$ are diagonal, local states amount to
diagonal density matrices
\begin{equation}
\rho_{{\cal E}}^{[0,n-1]}=\sum_{\bs{i}\in\Omega_D^n}
\mu_{\bs{i}}\,
|i_0\rangle\langle i_0|\otimes|i_1\rangle\langle i_1|\otimes\cdots
\otimes|i_{n-1}\rangle\langle i_{n-1}|\ ,
\label{SMaCSC_5}
\end{equation}
and the Shannon entropy~\eqref{KSE_2} to the Von Neumann entropy
\begin{equation}
S_\mu({\cal E}_{[0,n-1]})=-\Tr\left[\rho_{{\cal E}}^{[0,n-1]}\log
\rho_{{\cal E}}^{[0,n-1]}\right]\eqcol
H_\mu\left[{{\cal E}}_{[0,n-1]}\right]\ .
\label{SMaCSC_6}
\end{equation}
\section{ALF--Entropy}\label{AFE}
From an algebraic point of view, the difference between a triplet
$({\cal M},\omega, \Theta)$ describing a quantum dynamical system and
a triplet $\pt{{\cal A}_{\cal X}, \omega_\mu, \Theta_\alpha}$ as in
Definition~\ref{RoeiW_51c} is that $\omega$ and $\Theta$ are now a
$\Theta$--invariant state, respectively an automorphism over a
non--commutative (C* or Von Neumann) algebra of operators. 
\\[-2ex]
\begin{quote}
\begin{NNN}{}\ \\[-5ex]
\label{rem_41}
\begin{Ventry}{}
\item[] In finite dimension $D$, ${\cal M}$ is the full matrix algebra
of $D\times D$ matrices,
the states $\omega$ are given by density matrices 
$\rho_\omega$, such that 
\mbox{$\omega(X)\coleq\Tr(\rho_\omega X)$,}
while the reversible dynamics $\Theta$ is unitarily implemented: 
$\Theta(X)=UXU^*$.
\end{Ventry}
\end{NNN}
\end{quote}
\noindent
The quantum dynamical entropy proposed in~\cite{Ref_20} by Alicki and 
Fannes, ALF--entropy for short, is based on the idea that, in analogy
with what one does for the metric entropy, one can model symbolically
the evolution of quantum systems by means of the
right shift along a spin chain. In the quantum case the
finite--dimensional matrix algebras at the various sites are not
diagonal, but, typically, full matrix algebras, that is the spin at
each site is a quantum spin.

This is done by means of so called \enfasi{partitions of unit}, that
is by finite sets ${\cal Y}=\Big\{y_1,y_2,\ldots,y_D\Big\}$ of
operators in a $\Theta$--invariant subalgebra ${\cal M}_0\in {\cal M}$
such that
\begin{equation}
\sum_{\ell=1}^D y_\ell^*y_\ell^{\phantom{*}}=\Id\ ,
\label{AFE_1}
\end{equation}
where $y_j^*$ denotes the adjoint of $y_j^{\phantom{*}}$.
With ${\cal Y}$ and the state $\omega$ one constructs the $D\times D$ matrix
with entries $\omega(y_j^*y_i^{\phantom{*}})$;
such a matrix is a density matrix $\rho[{\cal Y}]$:
\begin{equation}
\rho{[{\cal Y}]}_{i,j}\coleq\omega(y_j^* y_i^{\phantom{*}})\ \cdot
\label{AFE_2}
\end{equation}
It is thus possible to define the entropy of a partition of unit as (compare~\eqref{SMaCSC_6}):
\begin{equation}
H_\omega[{\cal Y}]\coleq-\Tr\Big(\rho[{\cal Y}]\log\rho[{\cal Y}]\Big)\ \cdot
\label{AFE_3}
\end{equation}
Further, given two partitions of unit 
${\cal Y}=\Bigl(y_0,y_1,\ldots,y_D\Bigr)$,
${\cal Z}=\Bigl(z_0,z_1,\ldots,z_B\Bigr)$,
of size $D$, respectively $B$, one gets a finer partition  
of unit of size $BD$ as the set
\begin{equation}
\ \!\!\!\!\!{\cal Y}\circ {\cal Z}
\coleq\Big(
y_0 z_0,\cdots,y_0 z_B;
y_1 z_0,\cdots,y_1 z_B;\cdots;
y_D z_0,\cdots,y_D z_B \Big)\cdot
\label{AFE_4}
\end{equation}
After $j$ time--steps, ${\cal Y}$ evolves into  
$\Theta^j({\cal Y})\coleq\Big\{\Theta^j(y_1),\Theta^j(y_2),\cdots,
\Theta^j(y_D)\Big\}$. Since $\Theta$ is an automorphism,
$\Theta^j\pt{{\cal Y}}$ is a partition of unit;
then, one refines $\Theta^j({\cal Y})$, $0\leq j\leq n-1$,
into a larger partition of unit
\begin{alignat}{5}
{\cal Y}^{[0,n-1]}
& \coleq\Theta^{n-1}({\cal Y}) \ \; \circ
&\,
& \Theta^{n-2}({\cal Y}) \ \; \circ
&\,
& \cdots \ \; \circ \ \;
&\,
& \Theta({\cal Y})  \ \; \circ \ \; 
&\,
& {\cal Y}\cdot
\label{AFE_5}
\intertext{We shall denote the typical element of $\displaystyle
{\pq{{\cal Y}^{[0,n-1]}}}$ by}
{\pq{{\cal Y}^{[0,n-1]}}}_{\bs{i}} 
& = \Theta^{n-1}\pt{y_{i_{n-1}}^{\phantom{*}}}
&& \Theta^{n-2}\pt{y_{i_{n-2}}^{\phantom{*}}} 
&& \cdots 
&& \Theta(y_{i_1}^{\phantom{*}}) 
&& y_{i_0}^{\phantom{*}}\cdot
\label{AFE_55}
\end{alignat}
Each refinement is in turn associated with a density matrix
$\rho_{\cal Y}^{[0,n-1]}\coleq\rho\left[{\cal Y}^{[0,n-1]}\right]$
which is a state on the algebra 
$\displaystyle
{\bf M}_D^{[0,n-1]}\coleq\otimes_{\ell=0}^{n-1}{({\bf M}_D)}_\ell$,
with entries 
\begin{equation}
{\bigg[\rho\Big[{\cal Y}^{[0,n-1]}\Big]\bigg]}_{\bs{i},\bs{j}}\coleq
\omega\Big(
y_{j_0}^*\Theta\left(y_{j_1}^*\right)\cdots\Theta^{n-1}
\left(y_{j_{n-1}}^* y_{i_{n-1}}^{\phantom{*}}\right)
\cdots\Theta\left(y_{i_1}^{\phantom{*}}\right)y_{i_0}^{\phantom{*}}
\Big)\ \cdot
\label{AFE_6}
\end{equation}
Moreover each refinement has an entropy 
\begin{equation}
H_{\omega}\Big[{\cal Y}^{[0,n-1]}\Big] = -\Tr\Big(
\rho\Big[{\cal Y}^{[0,n-1]}\Big]\log
\rho\Big[{\cal Y}^{[0,n-1]}\Big]\Big)\ \cdot
\label{fabio3}
\end{equation}
The states $\rho_{\cal Y}^{[0,n-1]}$ are compatible%
: $\rho^{[0,n-1]}_{\cal Y}\rstr\;{\bf M}_D^{[0,n-2]}=
\rho^{[0,n-2]}_{\cal Y}$,
and define a global state
$\rho_{\cal Y}$ on the quantum spin chain 
$\displaystyle
{\bf M}_D^\infty\coleq\otimes_{\ell=0}^\infty({\bf M}_D)_\ell$.

Then, as in the previous Section, it is possible 
to associate with the quantum dynamical system $({\cal
M},\omega,\Theta)$ a symbolic dynamics which amounts to the
right--shift,
 $\displaystyle \sigma:{({\bf M}_D)}_\ell\mapsto{({\bf
M}_D)}_{\ell+1}$, along the quantum spin half--chain
(compare~\eqref{SMaCSC_3}).\\
Non--commutativity makes
$\rho_{\cal Y}$ not shift--invariant, in general~\cite{Ref_20}.
In this case, the existence of a limit as in~\eqref{KSE_3} is not
guaranteed and one has to define
the ALF--entropy of $({\cal M},\omega,\Theta)$ as 
\begin{subequations}
\label{AFE_7}
\begin{align}
h^{ALF}_{\omega,{\cal M}_0}(\Theta) &
\coleq\sup_{{\cal Y}\subset{\cal M}_0}
h^{ALF}_{\omega,{\cal M}_0}(\Theta,{\cal Y})\ ,
\label{AFE_7a} \\ 
\text{where}\qquad \qquad 
h^{ALF}_{\omega,{\cal M}_0}(\Theta,{\cal Y}) &  
\coleq\limsup_n \frac{1}{n} H_{\omega}\Big[{\cal Y}^{[0,n-1]}\Big]\ \cdot
\label{AFE_7b}
\end{align}  
\end{subequations}
Like the metric entropy of a partition ${\cal E}$, also the
ALF--entropy of a partition of unit ${\cal Y}$ can be physically
interpreted as an asymptotic \enfasi{entropy production} relative to a
specific coarse--graining.\\[-2ex]
\begin{quote}
\begin{NNN}{}\ \\[-0.5ex]
The ALF--entropy reduces to the Kolmogorov metric entropy on classical
systems. This is best seen by using an algebraic characterization of
$\tripC$ by means of the Von Neumann algebra ${\cal M}_{\cal X} =
\Lspace{\infty}{{\cal X}}$ of essentially bounded functions on ${\cal
X}$~\cite{Ref_28bis}. The characteristic functions of measurable
subsets of ${\cal X}$ constitute a *subalgebra ${\cal M}_0\subseteq
{\cal M}_{\cal X}$; moreover, given a partition $\cal E$ of $\cal X$,
the characteristic functions $e_{E_\ell}$ of its atoms $E_\ell$, 
$\displaystyle {\cal Z}_{{\cal E}}=\{e_{E_1},\cdots,e_{E_D}\}$
is a partition of unit in ${\cal M}_0$.
From~\eqref{AoDC_3} it follows that \mbox{$\displaystyle
\Theta_{\alpha}^j(e_{E_\ell})=e_{T_{\alpha}^{-j}(E_\ell)}$} and
from~\eqref{AoDC_2} that
$\displaystyle
{\Big[\rho\big[{\cal Z}_{{\cal E}}^{[0,n-1]}\big]\Big]}_{\bs{i},\bs{j}}
=\delta_{\bs{i},\bs{j}}\,\mu_{\bs{i}}$
(see~\eqref{KSE_1}),
whence $\displaystyle
H_\omega\big[{\cal Z}_{{\cal E}}^{[0,n-1]}\big]
=S_\mu\big({\cal E}_{[0,n-1]}\big)$
(see~\eqref{KSE_2} and~\eqref{AFE_3}). In such a case, the $\limsup$
in~\eqref{AFE_7b} is actually a true limit and yields~\eqref{KSE_3}.
In~\cite{Ref_30}, the same result is obtained by means of the algebra
${\cal A}_{\cal X}$ and of the *subalgebra ${\cal W}_{\text{exp}}$ of
exponential functions.
\end{NNN}
\end{quote}
\subsection{ALF--Entropy for Discretized $\tripC$}
We now return to the classical systems $({\cal A}_{\cal
X},\omega_\mu,\Theta_\alpha)$ of Section~\ref{AATT}.
For later use, we introduce the following map defined on 
the torus ${\IT}^2\pt{{[0,N)}^2}$, namely ${[0,N)}^2 \pmod{N}$, 
and on its subset ${\pt{\IZ / N \IZ}}^2$:
\begin{equation}
{\IT}^2\pt{{[0,N)}^2} \ni\bs{x}\mapsto U_\alpha\pt{\bs{x}} \coleq N\,
T_\alpha\pt{\frac{\bs{x}}{N}}\in{\IT}^2\pt{{[0,N)}^2}\label{Ualpha}
\end{equation}
\noindent
The use of the *morphisms ${\cal J}_{N ,
\infty}$ and ${\cal J}_{\infty, N}$ introduced in Section~\ref{roL},
makes it convenient to define the discretized versions of $({\cal A}_{\cal
X},\omega_\mu,\Theta_\alpha)$ as follows:\\[-2ex]
\begin{quote}
\begin{DDD}{}\ \\[-0.5ex]
A discretization of $({\cal A}_{\cal
X},\omega_\mu,\Theta_\alpha)$ is the triplet
$\pt{D_{N^2}\pt{\IC},\omega_{N^2},\widetilde{\Theta}_\alpha}$ where:
\begin{Ventry}{$D_{N^2}\pt{\IC}$}\label{RoeiW_51d}
	\item[$D_{N^2}\pt{\IC}$] is the abelian algebra of diagonal
	matrices acting on ${\IC}^{N^2}$.
	\item[$\omega_{N^2}$] is the {\it tracial state} given by the
	expectation: 
	\begin{equation}
	D_{N^2}\pt{\IC}\ni M \mapsto
	\omega_{N^2}\pt{M}\coleq\frac{1}{N^2}\;\Tr\pt{M}\label{CoAFE_2}. 
	\end{equation}
	\item[$\widetilde{\Theta}_\alpha$] is the *automorphism of
	$D_{N^2}\pt{\IC}$ defined by:
	\begin{equation}
	\!\!\!\!\!\!\!\!D_{N^2}\pt{\IC}\ni M \mapsto
	\widetilde{\Theta}_\alpha^{\phantom{t}}\pt{M}  
	\coleq\sum_{\bs{\ell} \in {(\ZNZ{N})^2}} 
	M_{U_\alpha\pt{\bs{\ell}},U_\alpha\pt{\bs{\ell}}}
	\ket{\bs{\ell}}\bra{\bs{\ell}}\
	\cdot \label{CoAFE_11}
	\end{equation}
\end{Ventry}
\end{DDD}
\end{quote}
\noindent
\\[-7.5ex]
\begin{quote}
\begin{NNS}{}\ \\[-5ex]\label{def_Vj}
\begin{Ventry}{\mdseries iii.}
\item[\mdseries i.] The expectation $\omega_{N^2}\pt{{\cal
J}_{N,\infty}\pt{f}}$
corresponds to the numerical calculation of the integral of $f$
 realized on a $N\times N$ grid on $\IT^2$. 
\item[\mdseries ii.] $\widetilde{\Theta}_\alpha^{\phantom{t}}$ is a
*automorphism because the map
${\pt{\IZ / N \IZ}}^2\ni\bs{\ell}\longmapsto U_\alpha\pt{\bs{\ell}}$ is a
bijection. For the same reason the state $\omega_{N^2}$ is $\widetilde{\Theta}_\alpha^{\phantom{t}}$--invariant.
\item[\mdseries iii.] One can check that, given $f\in{\cal A}_{\cal
X}$, 
\begin{equation}
\widetilde{\Theta}_\alpha^{\phantom{t}}\pt{{\cal J}_{N ,
\infty}\pt{f}} \coleq \sum_{\bs{\ell} \in {(\ZNZ{N})^2}} 
f\pt{\frac{U_\alpha\pt{\bs{\ell}}}{N}}
\ket{\bs{\ell}}\bra{\bs{\ell}}\ \cdot
\label{CoAFE_011}
\end{equation}
\item[\mdseries iv.]
Also, $\widetilde{\Theta}_\alpha^{j}\circ{\cal J}_{N ,
\infty}	= {\cal J}_{N ,\infty}\circ\Theta_\alpha^{j}$ for all $j\in\IZ$.
\item[\mdseries v.] On the contrary, for $j\in{\IN}$, 
$\widetilde{\Theta}_\alpha^{j}\circ{\cal J}_{N ,
\infty}\neq{\cal J}_{N ,\infty}\circ\Theta_\alpha^{j}$ for the
Sawtooth Maps, that is when $\alpha\not\in{\IZ}$.
\end{Ventry}
\end{NNS}
\end{quote}
The automorphism $\widetilde{\Theta}_\alpha^{\phantom{t}}$ can be
rewritten in the more familiar form
\begin{align}
\widetilde{\Theta}_\alpha^{\phantom{t}}\pt{X} & 
= \sum_{\bs{\ell} \in {(\ZNZ{N})^2}} 
X_{U_\alpha\pt{\bs{\ell}},U_\alpha\pt{\bs{\ell}}}
\ket{\bs{\ell}}\bra{\bs{\ell}}\notag\\
& = \sum_{U_\alpha^{-1}\pt{\bs{s}} \in {(\ZNZ{N})^2}} 
X_{\bs{s},\bs{s}}
\ket{U_\alpha^{-1}\pt{\bs{s}}}\bra{U_\alpha^{-1}{\bs{s}}}
\notag\\
\!\!\!\!\!\!\!\!\!{\textstyle\text{(see Remark~\ref{rem_44} i and ii)}}\qquad
& = U_{\alpha,N}^{\phantom{*}}\pt{\sum_{\newatop{\text{all equiv.}}{\text{classes}}} 
X_{\bs{s},\bs{s}}
\ket{\bs{s}}\bra{\bs{s}}}U_{\alpha,N}^{*} \label{CoAFE_121}\\
& = U_{\alpha,N}^{\phantom{*}}
\;X\;\;
U_{\alpha,N}^{*}\ ,\label{CoAFE_131}
\end{align}
where the operators $U_{\alpha,N}$ era defined by
\begin{equation}
{\cal H}_{N^2}\ni\big|\bs{\ell}\big\rangle\longmapsto
U_{\alpha,N}\big|\bs{\ell}\big\rangle\coleq\ket{
U_\alpha^{-1}\pt{\bs{\ell}}}\ \cdot\label{aggiunta}
\end{equation} 
\\[-7.5ex]
\begin{quote}
\begin{NNS}{}\ 
\label{rem_44}
\\[-5ex]
\begin{Ventry}{iii)}
	\item[\mdseries i)] All of $T_\alpha$, $T_\alpha^{-1}$,
	$T_\alpha^{t}$ and ${\pt{T_\alpha^{-1}}}^{t}$ belong to
	$SL_2\pt{\IZ/N\IZ}$; in particular these matrices are
	automorphisms on ${\pt{\IZ/N\IZ}}^2$ so that,
	in~\eqref{CoAFE_121}, one can sum over the equivalence classes.
	\item[\mdseries ii)] The same argument as before proves that
	the operators in~\eqref{aggiunta} are unitary which is
	equivalent to saying that $\widetilde{\Theta}_\alpha$ is a
	*automorphism.
\end{Ventry} 
\end{NNS}
\end{quote}
\noindent
In order to construct the ALF--entropy,
we now seek a useful partition of unit in 
$(D_{N^2},\omega_{N^2},\widetilde{\Theta}_\alpha)$; we do that by means 
of the subalgebra ${\cal
W}_\infty\subseteq{\cal A}_{\cal X}$ in
equations~(\ref{equ_22}--\ref{RoeiW_1}): 
\begin{align}
{\cal Y} \coleq{\bigg\{y_j\bigg\}}_{j = 1}^D  & =
{\pg{\frac{1}{\sqrt{D}}\;\exp\pt{2 \pi i \bs{r}_j\cdot\bs{x}}}}_{j =
1}^D\ , \label{CoAFE_21}\\
\text{where } \qquad {\bigg\{\bs{r}_j\bigg\}}_{j = 1}^D & \eqcol
\Lambda \subset 
{\pt{\IZ/N\IZ}}^2 \cdot \label{CoAFE_217}
\end{align}\\[-7.5ex]
\begin{quote}
\begin{DDD}{}\ \\[-0.5ex]\label{partizione}
Given a subset $\Lambda$ of the lattice consisting of the points
${\pg{\bs{r}_j}}$ as in~\eqref{CoAFE_217}, we shall denote by ${\cal
\widetilde{Y}}$ the partition of unit in
$(D_{N^2},\omega_{N^2},\widetilde{\Theta}_\alpha)$ given by:
\begin{equation}
{\cal \widetilde{Y}} = {\bigg\{\widetilde{y}_j\bigg\}}_{j = 1}^D \coleq
{\bigg\{{\cal J}_{N , \infty}\pt{y_j}\bigg\}}_{j = 1}^D =
{\pg{\frac{1}{\sqrt{D}}\;\widetilde{W}(\bs{r}_j)}}_{j = 1}^D\ ,
\label{CoAFE_3} 
\end{equation}
with $\widetilde{W}(\bs{r}_j)$ defined in~\eqref{RoeiW_4}.
\end{DDD}
\end{quote}
\noindent
From the above definition, the elements of the
refined partitions in~\eqref{AFE_55} take the form:
\begin{equation}
{\pq{{\cal \widetilde{Y}}^{[0,n-1]}}}_{\bs{i}} =
\frac{1}{N}\frac{1}{D^{\frac{n}{2}}}
\sum_{\bs{\ell} \in {(\ZNZ{N})^2}} 
e^{\frac{\:2\pi i}{N}\pq{
\bs{r}_{i_{n-1}}\cdot U_{\alpha}^{n-1} \pt{\bs{\ell}} + \,\cdots 
\,+\bs{r}_{i_1}\cdot U_{\alpha} \pt{\bs{\ell}} 
+ \bs{r}_{i_0}\cdot \bs{\ell}}}
\ket{\bs{\ell}}\bra{\bs{\ell}}\ \cdot
\label{CoAFE_31}
\end{equation}	
Then, the multitime correlation matrix $\rho_{\widetilde{\cal
Y}}^{\pq{0,n-1}}$ in~\eqref{AFE_6} has entries:
\begin{align}
{\bigg[\rho\Big[{\cal \widetilde Y}^{[0,n-1]}\Big]\bigg]}_{\bs{i},\bs{j}}
& = \frac{1}{N^2}\frac{1}{D^{n}}
\sum_{\bs{\ell} \in {(\ZNZ{N})^2}} 
e^{\frac{\:2\pi i }{N}
\overset{n-1}{\underset{p=0}{\sum}}
\pt{\bs{r}_{i_p}- \bs{r}_{j_p}} 
\cdot U_{\alpha}^{p} 
\pt{\bs{\ell}}}\ \ , \ \ U_{\alpha}^{0}\pt{\bs{\ell}} = {\Id}
\label{CoAFE_5}\\
& = \sum_{\bs{\ell} \in {(\ZNZ{N})^2}} 
\bkk{\bs{i}}{g_{\bs{\ell}}\pt{n}}\bkk{g_{\bs{\ell}}\pt{n}}{\bs{j}}\ , 
\label{CoAFE_6}\\
\text{with}\quad \bkk{\bs{i}}{g_{\bs{\ell}}\pt{n}}&\coleq
\frac{1}{N}\frac{1}{D^{\frac{n}{2}}}
e^{\frac{\:2\pi i }{N}
\overset{n-1}{\underset{p=0}{\sum}}
\bs{r}_{i_p} \cdot U_{\alpha}^{p} \pt{\bs{\ell}}}\in {\IC}^{D^n}\ .
\label{CoAFE_7}
\end{align}
The density matrix $\rho_{\widetilde{\cal
Y}}^{\pq{0,n-1}}$ can now be used
to numerically compute the ALF--entropy as in~\eqref{AFE_7};
however, the large dimension ($D^n\times D^n$) makes the computational
problem
very hard, a part for small numbers of
iterations. Our goal is to prove that another matrix (of fixed dimension $N^2
\times N^2$) can be used instead of $\rho_{\widetilde{\cal
Y}}^{\pq{0,n-1}}$.\\[-2ex]
\begin{quote}
\begin{PPP}{}\ \\[-5ex]
\begin{Ventry}{}\label{prop_41}
	\item[] Let ${\cal G}\pt{n}$ be the $N^2\times N^2$ matrix
	with entries
\begin{equation}
{\cal G}_{\bs{\ell}_1,\bs{\ell}_2}\pt{n} \coleq
\bkk{g_{\bs{\ell}_2}\pt{n}}{g_{\bs{\ell}_1}\pt{n}}
\label{CoAFE_1803}
\end{equation}
given by the scalar products of the vectors
$\ket{g_{\bs{\ell}}\pt{n}}\in {\cal H}_{D^n}=
{\IC}^{D^n}$ in~\eqref{CoAFE_7}.
Then, the entropy of the partition of unit ${\cal \widetilde
Y}^{[0,n-1]}$ with elements~\eqref{CoAFE_3} is given by:
\begin{equation}
H_{\omega_{N^2}}\pq{{\cal \widetilde
Y}^{[0,n-1]}}
= -{\Tr}_{{\cal H}_{N^2}}\Big({\cal G}\pt{n}\log{\cal G}\pt{n}\Big)
\label{CoAFE_10}
\end{equation}
\end{Ventry}
\end{PPP}
\end{quote}
\noindent 
{\bf Proof:}\quad ${\cal G}\pt{n}$ is hermitian and
from~\eqref{CoAFE_7} it follows that ${\Tr}_{{\cal H}_{N^2}}{\cal
G}\pt{n} =1$.\\
Let ${\cal H}_{N^2}=
{\IC}^{N^2},\ {\cal H}\coleq {\cal H}_{D^n}\otimes{\cal H}_{N^2}$ and
consider the projection $\rho_{\psi} =\ket{\psi}\bra{\psi}$ onto
\begin{equation}
{\cal H}\ni\ket{\psi} \coleq \sum_{\bs{\ell} \in {(\ZNZ{N})^2}}
\ket{g_{\bs{\ell}}\pt{n}}\otimes \ket{\bs{\ell}}.
\label{CoAFE_80}
\end{equation}
We denote by $\Sigma_1$ the restriction of $\rho_{\psi}$ to the full
matrix algebra $M_1\coleq M_{D^n}\pt{\IC}$ and by $\Sigma_2$ the
restriction to $M_2\coleq M_{N^2}\pt{\IC}$. It follows that:
\begin{equation*}
\Tr_{{\cal H}_{D^n}}\pt{\Sigma_1 \cdot m_1} = 
\bra{\psi} m_1\otimes{\Id}_2\ket{\psi}
= \sum_{\bs{\ell} \in {(\ZNZ{N})^2}}
\bra{g_{\bs{\ell}}} m_1
\ket{g_{\bs{\ell}}}\ , \ \forall m_1\in M_1 \ \cdot
\end{equation*}
Thus, from~\eqref{CoAFE_6},
\begin{equation}
\Sigma_1 = \rho_{\cal \widetilde{Y}}^{\pq{0,n-1}} = 
\sum_{\bs{\ell} \in {(\ZNZ{N})^2}} 
\ket{g_{\bs{\ell}}\pt{n}}\bra{g_{\bs{\ell}}\pt{n}}\
\cdot\label{CoAFE_801}
\end{equation}
On the other hand, from
\begin{align*}
\Tr_{{\cal H}_{N^2}}\pt{\Sigma_2 \cdot m_2} & = 
\bra{\psi} {\Id}_1\otimes m_2\ket{\psi}\\
& = 
\sum_{\bs{\ell}_1,\bs{\ell}_2 \in {(\ZNZ{N})^2}}
\bkk{g_{\bs{\ell}_2}\pt{n}}{g_{\bs{\ell}_1}\pt{n}}
\langle\bs{\ell}_2
| m_2 |
\bs{\ell}_1\rangle
\ , \ \forall m_2\in M_2\ ,
\end{align*}
it turns out that $\Sigma_2 = {\cal G}\pt{n}$, whence the result
follows from Araki--Lieb's inequality~\cite{Ref_29}.\hfill$\qed$

\noindent We now return to the explicit computation of the density
matrix ${\cal G}\pt{n}$ in Proposition~\ref{prop_41}.
By using the transposed matrix $T_\alpha^{\text{tr}}$, the
vectors~\eqref{CoAFE_7} now read
\begin{align}
\bkk{\bs{i}}{g_{\bs{\ell}}\pt{n}} & =
\frac{1}{N\,D^{\frac{n}{2}}}
e^{\frac{\:2\pi i}{N}\;\bs{\ell}\cdot
\bs{f}_{\Lambda,\alpha}^{(n),N}\pt{\bs{i}}}
\label{CoAFE_710}\\
\bs{f}_{\Lambda,\alpha}^{(n),N}\pt{\bs{i}} & \coleq
\sum_{p=0}^{n-1}\;{\pt{T_\alpha^{\text{tr}}}}^{p} \,\bs{r}_{i_p}
\pmod{N} \label{CoAFE_720}
\end{align}
where we made explicit the various dependencies of~\eqref{CoAFE_720}
on $n$ the time--step, $N$ the inverse lattice--spacing, the chosen set
$\Lambda$ of $\bs{r}_j$'s and the $\alpha$ parameter of the dynamics
in ${\text{SL}}_2 {(\ZNZ{N})^2}$.

\noindent In the following we shall use the equivalence classes 
\begin{equation}
\pq{\bs{r}}  \coleq\pg{\bs{i}\in\Omega_{D}^{(n)} \;\Big|\; 
\bs{f}_{\Lambda,\alpha}^{(n),N}
\pt{\bs{i}}\equiv\bs{r}\in{(\ZNZ{N})^2}\pmod{N}}\ ,
\label{CoAFE_730}
\end{equation}
their
cardinalities $\# \pq{\bs{r}}$ and, in particular, the frequency
function $\nu_{\Lambda,\alpha}^{(n),N}$
\begin{equation}
\ZNZD\ni\bs{r} \longmapsto\nu_{\Lambda,\alpha}^{(n),N}\pt{\bs{r}}
\coleq 
\frac{\# \pq{\bs{r}}}{D^n}\ \cdot
\label{CoAFE_740}
\end{equation}
\\[-7.5ex]

\begin{quote}
\begin{PPP}{}\ \\[-5ex]
\begin{Ventry}{}\label{prop_42}
	\item[] The Von Neumann entropy of the refined (exponential)
	partition of unit up to time $n-1$ is given by:
\begin{equation}
H_{\omega_{N^2}}\pq{{\cal \widetilde Y}^{[0,n-1]}} =
- \sum_{\bs{r} \in {(\ZNZ{N})^2}} 
\nu_{\Lambda,\alpha}^{(n),N}\pt{\bs{r}}
\log\, \nu_{\Lambda,\alpha}^{(n),N}\pt{\bs{r}}\label{CoAFE_77}
\end{equation}
\end{Ventry}
\end{PPP}
\end{quote}
\noindent
{\bf Proof:}\quad Using ~\eqref{CoAFE_710}, the matrix ${\cal
G}\pt{n}$ in Proposition~\ref{prop_41} can be written as:
\begin{align}
{\cal G}\pt{n} & = 
\frac{1}{D^n}\sum_{\bs{i}\in\Omega_D^n} 
\ket{f_{\bs{i}}\pt{n}}\bra{f_{\bs{i}}\pt{n}}\ ,
\label{CoAFE_421}\\
\bkk{\,\bs{\ell}\,}{f_{\bs{i}}\pt{n}} & =
\frac{1}{N}
e^{\frac{\:2\pi i}{N}\;
\bs{f}_{\Lambda,\alpha}^{(n),N}\pt{\bs{i}}\cdot\bs{\ell}}
\label{CoAFE_422}
\end{align}
The vectors $\ket{f_{\bs{i}}\pt{n}}\in {\cal H}_{N^2} = {\IC}^{N^2}$
are such that
$\displaystyle
\bkk{f_{\bs{i}}\pt{n}}{f_{\bs{j}}\pt{n}} = 
\delta_{
\bs{f}_{\Lambda,\alpha}^{(n),N}\pt{\bs{i}}
\;,\;
\bs{f}_{\Lambda,\alpha}^{(n),N}\pt{\bs{j}}
}^{\pt{N}}$,
where with $\delta^{\pt{N}}$ is the $N$--periodic Kronecker delta.
For sake of simplicity, we say that $\ket{f_{\bs{i}}\pt{n}}$ belongs
to the equivalence class $\pq{\bs{r}}$ in~\eqref{CoAFE_730} if
$\bs{i}\in\pq{\bs{r}}$; vectors in different equivalence classes are
thus orthogonal, whereas those in a same equivalence class
$\pq{\bs{r}}$ are such that
\begin{align}
\bra{\bs{\ell}_1}
\pt{\sum_{\bs{i}\in \pq{\bs{r}}} 
\ket{f_{\bs{i}}\pt{n}}\bra{f_{\bs{i}}\pt{n}}}
\ket{\bs{\ell}_2} & = \frac{1}{N^2}
\sum_{\bs{i}\in \pq{\bs{r}}}
e^{\frac{\:2\pi i}{N}\;
\bs{f}_{\Lambda,\alpha}^{(n),N}\pt{\bs{i}}\cdot
\pt{\bs{\ell}_1-\bs{\ell}_2}}\notag\\
& = D^n \, \nu_{\Lambda,\alpha}^{(n),N}\pt{\bs{r}} 
\;\bkk{\bs{\ell}_1}{\bs{e}\pt{\bs{r}}}
\bkk{\bs{e}\pt{\bs{r}}}{\bs{\ell}_2}\notag\\
\bkk{\bs{\ell}}{\bs{e}\pt{\bs{r}}} & = \frac{e^{
\frac{\:2\pi i}{N}\;\bs{r}\cdot\bs{\ell}}}{N}
\in {\cal H}_{N^2}\notag
\end{align}
Therefore, the result follows from the spectral decomposition\\[2ex]  
$\phantom{a} $\hspace{36mm}$\displaystyle
{\cal G}\pt{n} = 
\sum_{\bs{r} \in {(\ZNZ{N})^2}} 
\nu_{\Lambda,\alpha}^{(n),N}\pt{\bs{r}}
\ket{\bs{e}\pt{\bs{r}}}\bra{\bs{e}\pt{\bs{r}}}\cdot\hfill\qed$
\section{Analysis of entropy production} 
In agreement with the intuition that finitely many states cannot
sustain any lasting entropy production, the ALF--entropy is indeed
zero for such systems~\cite{Ref_20}. However, this does not mean that
the dynamics may not be able to show a significant entropy rate over
finite interval of times, these being typical of the underlying
dynamics.

As already observed in the Introduction, in quantum chaos one deals
with quantized classically chaotic systems; there, one finds that
classical and quantum mechanics are both correct descriptions
over times scaling
with $\log\hbar^{-1}$. Therefore, the classical--quantum
correspondence occurs over times much smaller than the 
Heisenberg recursion time that typically scales as $\hbar^{-\alpha},\
\alpha>0$.  
In other words, for quantized classically chaotic systems, the
classical description has to be replaced by the quantum one much
sooner than for integrable systems.

In this paper, we are considering not the quantization of
classical systems, but their discretization; nevertheless, we have
seen that, under certain respects, quantization and discretization are
like procedures with the inverse of the number of states $N$ playing
the role of $\hbar$ in the latter case.

We are then interested to study how the classical continuous behaviour
emerges from the discretized one when $N\to\infty$; in particular,
we want to investigate the presence of characteristic time scales and
of ``breaking--times'' $\tau_B$, namely those times beyond which the
discretized systems cease to produce entropy because their granularity
takes over and the dynamics reveals in full its regularity. 

Propositions~\ref{prop_41} and~\ref{prop_42} afford useful means to attack
such a problem numerically. In the following we shall be concerned
with the time behavior of the 
entropy of partition of units as in Definition~\ref{partizione}, the
presence of breaking--times $\tau_B\pt{\Lambda,N,\alpha}$, and their
dependence 
on the set $\Lambda$, on the number of states $N$ and on the dynamical
parameter $\alpha$.

As we shall see, in many cases $\tau_B$ depends quite 
heavily on the chosen partition of unit; we shall then try to cook up
a strategy to find a $\tau_B$ as stable as possible upon variation of
partitions, being led by the idea that the ``true'' $\tau_B$ has to be
strongly related to the Lyapounov exponent of the underlying
continuous dynamical system.

Equations~\eqref{CoAFE_10} and~\eqref{CoAFE_77} allow us to compute
the Von Neumann entropy of the state $\rho_{\cal
\widetilde{Y}}^{\pq{0,n-1}}$; 
if we were to compute
the ALF--entropy according to the definitions~\eqref{AFE_7}, the
result would be
zero, in agreement with fact that
the Lyapounov exponent for a
system with a finite number of states vanishes. 
Indeed, it is sufficient to notice that
the entropy $H_{\omega_{N^2}}\pq{{\cal \widetilde
Y}^{[0,n-1]}}$ is bounded from above by the entropy of the tracial
state $\displaystyle \frac{1}{N^2}{\Id}_{N^2}$, that is by $2\log N$;
therefore the expression 
\begin{equation}
h_{\omega_{N^2},{\cal W}_\infty}(\alpha,\Lambda,n)  
\coleq\frac{1}{n} H_{\omega_{N^2}}\Big[\widetilde{\cal Y}^{[0,n-1]}\Big],
\label{AEP_1}
\end{equation}
goes to zero with $n\longrightarrow 0$. It is for this reason that, in
the following, we will focus upon the temporal evolution of the
function $h_{\omega_{N^2},{\cal W}_\infty}(\alpha,\Lambda,n)$
instead of taking its $\limsup$ over the number of iterations $n$.

In the same spirit, we will not take the supremum of~\eqref{AEP_1}
over all possible partitions $\widetilde{\cal Y}$ (originated by
different $\Lambda$); instead, we will study 
the dependence of 
$h_{\omega_{N^2},{\cal W}_\infty}(\alpha,\Lambda,n)$
on different choices of partitions. 
In fact, if we vary over all possible choices of partitions of unit, we could
choose $\Lambda=\ZNZD$ in ~(\ref{CoAFE_217}), that is $D=N^2$;
then summation over all possible  
$\bs{r}\in\ZNZD$ would make the matrix elements 
$\displaystyle {\cal G}_{\bs{\ell}_1,\bs{\ell}_2}\pt{n}$ 
in~(\ref{CoAFE_1803}) equal to 
$\displaystyle \frac{\delta_{\bs{\ell_1},\bs{\ell_2}}}{N^2}$, whence 
$H_{\omega_{N^2}}\Big[\widetilde{\cal Y}^{[0,n-1]}\Big]=2\log N$.

\subsection{The case of
$T_{\alpha}\in\text{GL}_2\left(\IT^2\right)$} 
The maximum of $H_{\omega_{N^2}}$ is reached when the
frequencies~\eqref{CoAFE_740}
\begin{equation*}
\nu_{\Lambda,\alpha}^{(n),N}:{(\ZNZ{N})^2}\mapsto\pq{0,1}
\end{equation*}
become equal to $1/N^2$ over the torus: we will see that
this is indeed what happens to the frequencies
$\nu_{\Lambda,\alpha}^{(n),N}$ with $n\longrightarrow\infty$.
The latter behaviour
can be reached in various ways depending on:
\begin{itemize}
\item hyperbolic or elliptic regimes, namely on the dynamical
parameter $\alpha$;
\item number of elements ($D$) in the partition $\Lambda$;
\item mutual location of the $D$ elements $\bs{r}_i$ in $\Lambda$.
\end{itemize}
For later use we introduce the set of grid points with non--zero
frequencies 
\begin{equation}
\Gamma_{\Lambda,\alpha}^{(n),N} \coleq\pg{\frac{\bs{\ell}}{N}\ \bigg|\
\bs{\ell}\in{\pt{\IZ/N\IZ}}^2\ ,\ 
\nu_{\Lambda,\alpha}^{(n),N}\pt{\bs{\ell}}\neq 0}\ \cdot\label{gamma}
\end{equation}
\subsubsection{Hyperbolic regime with  $D$ randomly
chosen points $\bs{r}_i$ in $\Lambda$}\label{sez_511}
In the hyperbolic regime corresponding to $\alpha\in{\IZ}\setminus\pg{-4,-3,-2,-1,\phantom{-}0}$, $\Gamma_{\Lambda,\alpha}^{(n),N}$ tends to
increase its cardinality with the number of time--steps $n$.
Roughly speaking, there appear to be two distinct temporal patterns:
a first one, during which
\mbox{$\#\pt{\Gamma_{\Lambda,\alpha}^{(n),N}} \simeq D^n\leq N^2$} and
almost every 
$\nu_{\Lambda,\alpha}^{(n),N}\simeq D^{-n}$, followed by a second one
characterized by frequencies frozen to
$\nu_{\Lambda,\alpha}^{(n),N}\pt{\bs{\ell}} = \frac{1}{N^2}, 
\ \forall \bs{\ell}\in{(\ZNZ{N})^2}$.
The second temporal pattern is reached when, during the first one,
$\Gamma_{\Lambda,\alpha}^{(n),N}$ has covered the whole lattice and
$D^n\simeq N^2$.

From the point of view of the entropies, the first temporal regime is
characterized by
\begin{alignat}{2}
H_{\omega_{N^2}}(\alpha,\Lambda,n)  & \sim n\cdot \log D\qquad & ,\qquad
h_{\omega_{N^2},{\cal W}_\infty}(\alpha,\Lambda,n)  & \sim \log D\ ,\nonumber
\intertext{while the second one by}
H_{\omega_{N^2}}(\alpha,\Lambda,n)  & \sim 2 \log N\nonumber\qquad & ,\qquad
h_{\omega_{N^2},{\cal W}_\infty}(\alpha,\Lambda,n)  & \sim \frac{2 \log
N}{n}\ \cdot \nonumber
\end{alignat}
The transition between these two regimes occurs at $\bar{n} =
{\log}_D N^2$. However this time cannot be considered a realistic
breaking--time, as it too strongly depends on the chosen
partition.

Figure~\ref{lontani} (columns a and c) shows the mechanism clearly
in a density plot: white or light--grey points correspond to points of
$\Gamma_{\Lambda,\alpha}^{(n),N}$ and their number increases for small
numbers of iterations until the plot assume a \mbox{uniform grey color for
large $n$.} 

\begin{figure}[h]
\begin{center}
\ \\[0.7ex]
\begin{picture}(130,70)(0,0) 
\put(10,80){\epsfig{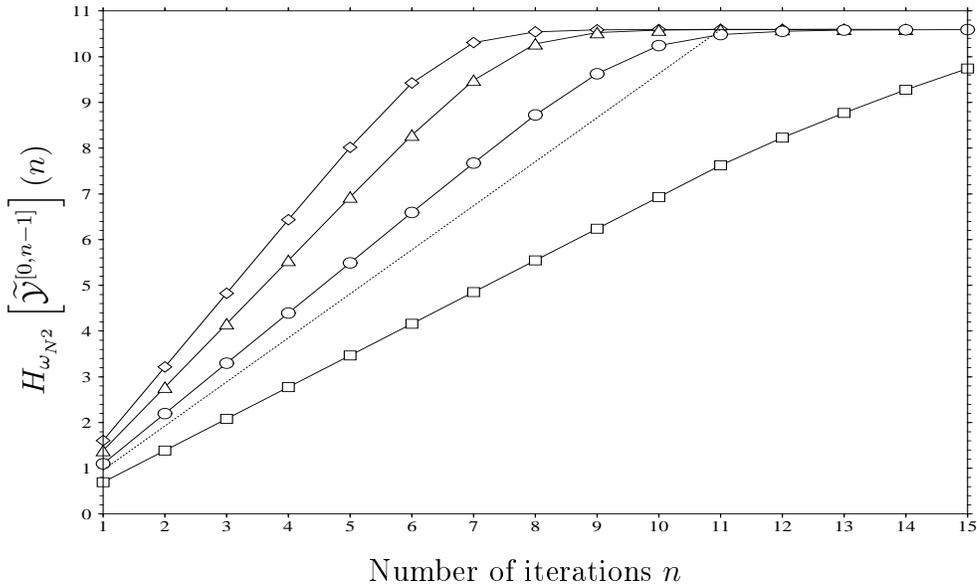}}
\begin{sideways}
\put(10,0){\makebox(70,10)[cc]{$\displaystyle H_{\omega_{N^2}}\pq{{\cal \widetilde
Y}^{[0,n-1]}}\pt{n}$}}
\end{sideways}
\put(0,0){\makebox(120,10)[cc]{Number of iterations $n$}}
\end{picture}
\caption{Von Neumann entropy $\displaystyle H_{\omega_{N^2}}\pt{n}$ 
in four hyperbolic ($\alpha = 1$) cases, for $D$
randomly distributed $\bs{r}_i$ in 
$\Lambda$, with $N=200$. Value for $D$ are: $\diamond = 5$,
\mbox{{\scriptsize 
$\bigtriangleup$} $= 4$}, $\circ = 3$ and {\scriptsize $\Box$} $=
2$. The dotted line represents $\displaystyle H_{\omega_{N^2}}\pt{n} =
\log\lambda \cdot n$ where $\log\lambda =
0.962\ldots$ is the Lyapounov exponent at$\alpha = 1$.}
\label{due}
\end{center}
\end{figure}

The linear and stationary behaviors of
$H_{\omega_{N^2}}(\alpha,\Lambda,n)$ are apparent 
in fig.~\ref{uno}, where four different plateaus
($2\log N $) are reached for four different $N$, and in fig.~\ref{due},
in which four different slopes are showed for four different number of
elements in the partition. With the same parameters as in
fig.~\ref{due}, fig.~\ref{tre} shows the corresponding
entropy production $h_{\omega_{N^2},{\cal W}_\infty}(\alpha,\Lambda,n)$.
\subsubsection{Hyperbolic regime with  $D$ nearest neighbors
$\bs{r}_i$ in $\Lambda$}\label{sub_512}
\begin{figure}[h]
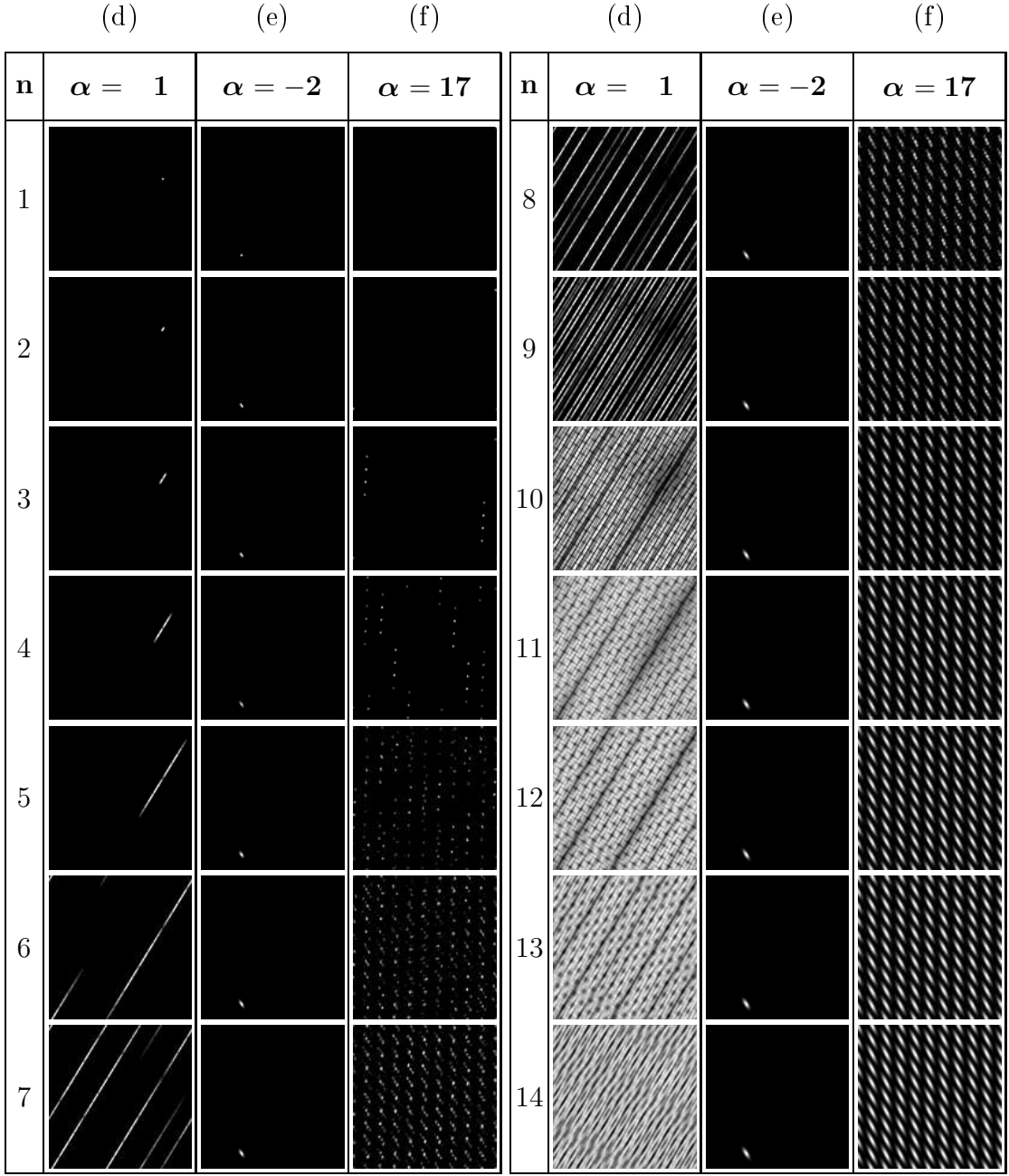

\begin{center}
\begin{picture}(140,164)(0,0) 
\put(0,132){\epsfig{width=\fotogram,file=d_01.epsi}
\epsfig{width=\fotogram,file=e_01.epsi}
\epsfig{width=\fotogram,file=f_01.epsi}\hspace{\sepfoto}
\epsfig{width=\fotogram,file=d_08.epsi}
\epsfig{width=\fotogram,file=e_08.epsi}
\epsfig{width=\fotogram,file=f_08.epsi}}
\put(0,110){\epsfig{width=\fotogram,file=d_02.epsi}
\epsfig{width=\fotogram,file=e_02.epsi}
\epsfig{width=\fotogram,file=f_02.epsi}\hspace{\sepfoto}
\epsfig{width=\fotogram,file=d_09.epsi}
\epsfig{width=\fotogram,file=e_09.epsi}
\epsfig{width=\fotogram,file=f_09.epsi}}
\put(0,88){\epsfig{width=\fotogram,file=d_03.epsi}
\epsfig{width=\fotogram,file=e_03.epsi}
\epsfig{width=\fotogram,file=f_03.epsi}\hspace{\sepfoto}
\epsfig{width=\fotogram,file=d_10.epsi}
\epsfig{width=\fotogram,file=e_10.epsi}
\epsfig{width=\fotogram,file=f_10.epsi}}
\put(0,66){\epsfig{width=\fotogram,file=d_04.epsi}
\epsfig{width=\fotogram,file=e_04.epsi}
\epsfig{width=\fotogram,file=f_04.epsi}\hspace{\sepfoto}
\epsfig{width=\fotogram,file=d_11.epsi}
\epsfig{width=\fotogram,file=e_11.epsi}
\epsfig{width=\fotogram,file=f_11.epsi}}
\put(0,44){\epsfig{width=\fotogram,file=d_05.epsi}
\epsfig{width=\fotogram,file=e_05.epsi}
\epsfig{width=\fotogram,file=f_05.epsi}\hspace{\sepfoto}
\epsfig{width=\fotogram,file=d_12.epsi}
\epsfig{width=\fotogram,file=e_12.epsi}
\epsfig{width=\fotogram,file=f_12.epsi}}
\put(0,22){\epsfig{width=\fotogram,file=d_06.epsi}
\epsfig{width=\fotogram,file=e_06.epsi}
\epsfig{width=\fotogram,file=f_06.epsi}\hspace{\sepfoto}
\epsfig{width=\fotogram,file=d_13.epsi}
\epsfig{width=\fotogram,file=e_13.epsi}
\epsfig{width=\fotogram,file=f_13.epsi}}
\put(0,0){\epsfig{width=\fotogram,file=d_07.epsi}
\epsfig{width=\fotogram,file=e_07.epsi}
\epsfig{width=\fotogram,file=f_07.epsi}\hspace{\sepfoto}
\epsfig{width=\fotogram,file=d_14.epsi}
\epsfig{width=\fotogram,file=e_14.epsi}
\epsfig{width=\fotogram,file=f_14.epsi}}

\put(-6.42,164){\line(0,-6){164.7}}
\put(-0.82,164){\line(0,-6){164.7}}
\put(21.58,164){\line(0,-6){164.7}}
\put(43.98,164){\line(0,-6){164.7}}
\put(66.38,164){\line(0,-6){164.7}}
\put(67.78,164){\line(0,-6){164.7}}
\put(73.38,164){\line(0,-6){164.7}}
\put(95.78,164){\line(0,-6){164.7}}
\put(118.18,164){\line(0,-6){164.7}}
\put(140.58,164){\line(0,-6){164.7}}
\put(-6.48,164){\line(6,0){72.94}}
\put(-6.48,154){\line(6,0){72.94}}
\put(-6.48,-0.7){\line(6,0){72.94}}
\put(67.72,164){\line(6,0){72.94}}
\put(67.72,154){\line(6,0){72.94}}
\put(67.72,-0.7){\line(6,0){72.94}}
\put(-6.42,154){\makebox(5.6,10)[cc]{$\mathbf{n}$}}
\put(-0.82,154){\makebox(22.4,10)[cc]{$\boldsymbol{\alpha}\ \mathbf{= \phantom{-}1}$}}
\put(21.58,154){\makebox(22.4,10)[cc]{$\boldsymbol{\alpha}\ \mathbf{= -2}$}}
\put(43.98,154){\makebox(22.4,10)[cc]{$\boldsymbol{\alpha}\ \mathbf{= 17}$}}
\put(-0.82,164){\makebox(22.4,10)[cc]{(d)}}
\put(21.58,164){\makebox(22.4,10)[cc]{(e)}}
\put(43.98,164){\makebox(22.4,10)[cc]{(f)}}
\put(67.78,154){\makebox(5.6,10)[cc]{$\mathbf{n}$}}
\put(73.38,154){\makebox(22.4,10)[cc]{$\boldsymbol{\alpha}\ \mathbf{= \phantom{-}1}$}}
\put(95.78,154){\makebox(22.4,10)[cc]{$\boldsymbol{\alpha}\ \mathbf{= -2}$}}
\put(118.18,154){\makebox(22.4,10)[cc]{$\boldsymbol{\alpha}\ \mathbf{= 17}$}}
\put(73.38,164){\makebox(22.4,10)[cc]{(d)}}
\put(95.78,164){\makebox(22.4,10)[cc]{(e)}}
\put(118.18,164){\makebox(22.4,10)[cc]{(f)}}
\put(-6.42,131.5){\makebox(5.6,22)[cc]{$1$}}
\put(-6.42,109.5){\makebox(5.6,22)[cc]{$2$}}
\put(-6.42,87.5){\makebox(5.6,22)[cc]{$3$}}
\put(-6.42,65.5){\makebox(5.6,22)[cc]{$4$}}
\put(-6.42,43.5){\makebox(5.6,22)[cc]{$5$}}
\put(-6.42,21.5){\makebox(5.6,22)[cc]{$6$}}
\put(-6.42,-0.5){\makebox(5.6,22)[cc]{$7$}}
\put(67.78,131.5){\makebox(5.6,22)[cc]{$8$}}
\put(67.78,109.5){\makebox(5.6,22)[cc]{$9$}}
\put(67.78,87.5){\makebox(5.6,22)[cc]{$10$}}
\put(67.78,65.5){\makebox(5.6,22)[cc]{$11$}}
\put(67.78,43.5){\makebox(5.6,22)[cc]{$12$}}
\put(67.78,21.5){\makebox(5.6,22)[cc]{$13$}}
\put(67.78,-0.5){\makebox(5.6,22)[cc]{$14$}}
\end{picture}
\caption{Density plots showing 
$\displaystyle \nu_{\Lambda,\alpha}^{(n),N}$
in two hyperbolic (columns d and f) and
one elliptic (col. e) regime, for five nearest neighboring $\bs{r}_i$
in $\Lambda$ ($N=200$). Black corresponds to 
$\displaystyle \nu_{\Lambda,\alpha}^{(n),N} = 0$.
When the system is chaotic, the frequencies tend to equidistribute on
${\pt{\IZ/N\IZ}}^2$ with
increasing $n$ and to approach, when the
breaking--time is reached,
the constant value $\frac{1}{N^2}$. Col. (f) shows how the dynamics
can be confined on a sublattice by a particular combination
$(\alpha,N,\Lambda)$ with a corresponding entropy decrease.}
\label{vicini}
\end{center}
\end{figure}

\begin{figure}[h]
\begin{center}
\ \\[0.2ex]
\begin{picture}(130,70)(0,0) 
\put(10,80){\epsfig{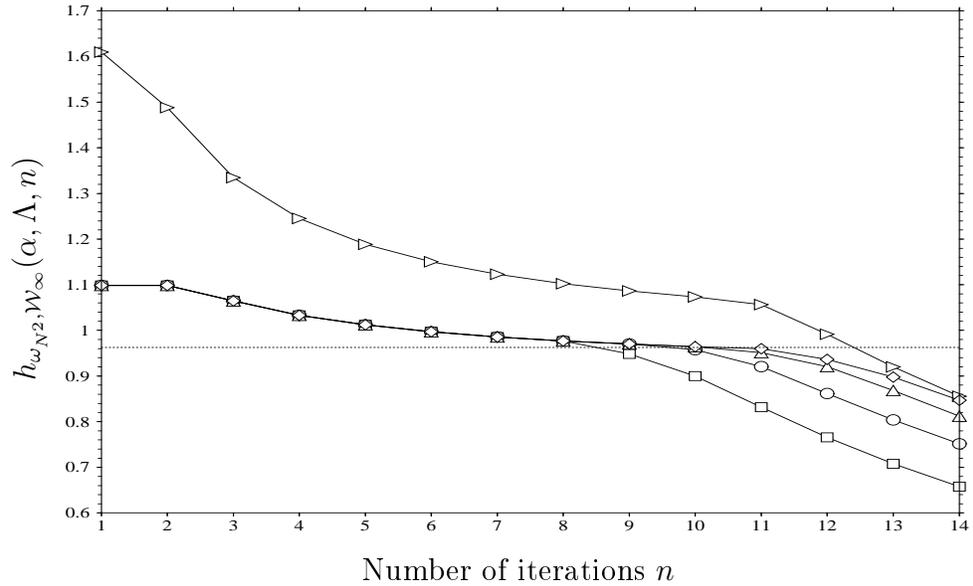}}
\begin{sideways}
\put(10,0){\makebox(70,10)[cc]{$\displaystyle h_{\omega_{N^2},{\cal W}_\infty}(\alpha,\Lambda,n)$}}
\end{sideways}
\put(0,0){\makebox(120,10)[cc]{Number of iterations $n$}}
\end{picture}
\caption{ Entropy production $\displaystyle h_{\omega_{N^2},{\cal
W}_\infty}(\alpha,\Lambda,n)$  in five hyperbolic
($\alpha = 1$) cases, for $D$ nearest neighboring points $\bs{r}_i$
in $\Lambda$. Values for $\pt{N,D}$ are: $\triangleright= \pt{200,5}$,
$\diamond = \pt{500,3}$, \mbox{{\scriptsize 
$\bigtriangleup$} $= \pt{400,3}$}, $\circ = \pt{300,3}$ and
{\scriptsize $\Box$} $= \pt{200,3}$. The dotted line corresponds to
the Lyapounov exponent $\log\lambda = 0.962\ldots$ at
$\alpha = 1$ and represents the natural asymptote for all these curves
in absence of breaking--time.} 
\label{cinque}
\end{center}
\end{figure}

In the following, we will consider a set of points $\Lambda =
{\pg{\bs{r}_i}}_{i=1\ldots D}$ very close to each other, instances of
which are as below:

%
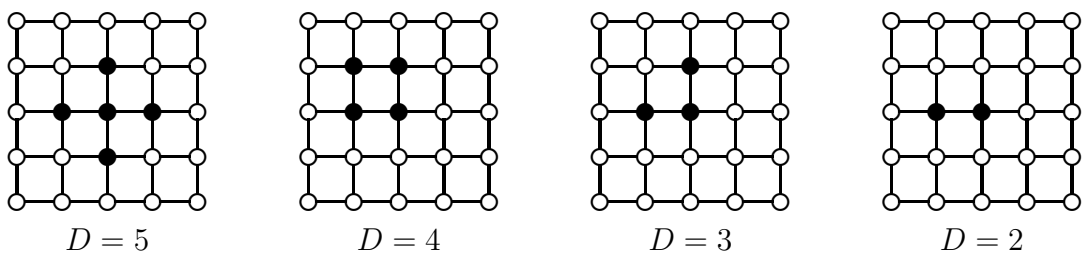
\begin{figure}[h]
\begin{center}
\begin{picture}(26,34)(-1,-9) 
\thicklines
\matrixput(0,0)(6,0){5}(0,6){5}{\circle{2}}
\matrixput(1,0)(6,0){4}(0,6){5}{\line(1,0){4}}
\matrixput(0,1)(6,0){5}(0,6){4}{\line(0,1){4}}
\put(12,12){\circle*{2}}
\put(6,12){\circle*{2}}
\put(12,6){\circle*{2}}
\put(12,18){\circle*{2}}
\put(18,12){\circle*{2}}
\put(0,-9){\makebox(24,8)[cc]{$D=5$}}
\end{picture}
\hspace{10mm}
\begin{picture}(26,34)(-1,-9) 
\thicklines
\matrixput(0,0)(6,0){5}(0,6){5}{\circle{2}}
\matrixput(1,0)(6,0){4}(0,6){5}{\line(1,0){4}}
\matrixput(0,1)(6,0){5}(0,6){4}{\line(0,1){4}}
\put(12,12){\circle*{2}}
\put(6,12){\circle*{2}}
\put(12,18){\circle*{2}}
\put(6,18){\circle*{2}}
\put(0,-9){\makebox(24,8)[cc]{$D=4$}}
\end{picture}
\hspace{10mm}
\begin{picture}(26,34)(-1,-9) 
\thicklines
\matrixput(0,0)(6,0){5}(0,6){5}{\circle{2}}
\matrixput(1,0)(6,0){4}(0,6){5}{\line(1,0){4}}
\matrixput(0,1)(6,0){5}(0,6){4}{\line(0,1){4}}
\put(12,12){\circle*{2}}
\put(6,12){\circle*{2}}
\put(12,18){\circle*{2}}
\put(0,-9){\makebox(24,8)[cc]{$D=3$}}
\end{picture}
\hspace{10mm}
\begin{picture}(26,34)(-1,-9) 
\thicklines
\matrixput(0,0)(6,0){5}(0,6){5}{\circle{2}}
\matrixput(1,0)(6,0){4}(0,6){5}{\line(1,0){4}}
\matrixput(0,1)(6,0){5}(0,6){4}{\line(0,1){4}}
\put(12,12){\circle*{2}}
\put(6,12){\circle*{2}}
\put(0,-9){\makebox(24,8)[cc]{$D=2$}}
\end{picture}
\caption{Several combinations of $D$ nearest neighbors in $\Lambda$
for different values $D$.}
\label{lattice}
\end{center}
\end{figure}
From eqs.~(\ref{CoAFE_730}--\ref{CoAFE_740}), the frequencies
$\nu_{\Lambda,\alpha}^{(n),N}\pt{\bs{\ell}}$ 
result proportional to 
how many strings have equal images $\bs{\ell}$, through the function
$\bs{f}_{\Lambda,\alpha}^{(n),N}$ in~\eqref{CoAFE_720}.
Due to the fact that
${\pq{T_\alpha}}_{11}={\pq{T_\alpha}}_{21}=1$, non--injectivity of
$\bs{f}_{\Lambda,\alpha}^{(n),N}$ occurs very frequently when
${\pg{\bs{r}_i}}$ are very close to each other. This is a dynamical
effect that, in continuous systems~\cite{Ref_30}, leads to an entropy
production  
approaching the Lyapounov exponent. Even in the discrete case, during
a finite time interval though, 
$h_{\omega_{N^2},{\cal W}_\infty}(\alpha,\Lambda,n)$ exhibits the same
behavior until $H_{\omega_{N^2}}$ reaches the upper bound $2\log N$.
From then on, the system behaves as described in
subsection~\ref{sez_511}, and the entropy production goes to zero as:
\begin{equation*}
h_{\omega_{N^2},{\cal W}_\infty}(\alpha,\Lambda,n) \sim
\frac{1}{n}\quad\text{(see fig.~\ref{cinque}).}
\end{equation*}
Concerning figure~\ref{vicini} (column d), 
whose corresponding graph
for $h_{\omega_{N^2},{\cal W}_\infty}(1,\Lambda,n)$ is 
labeled by $\triangleright$ in 
fig.~\ref{cinque}, we make the following consideration:
\begin{itemize}
\item for $n=1$ the white spot corresponds to five $\bs{r}_i$ grouped
as in fig.~\ref{lattice}.\\
In this case $h_{\omega_{N^2},{\cal
W}_\infty}(1,\Lambda,1)=\log D=\log 5$;
\item for $n\in \pq{2,5}$ the white spot begins to stretch along the
stretching direction of $T_1$. In this case,
the frequencies $\nu_{\Lambda,\alpha}^{(n),N}$ are not constant on the
light--grey region: this leads to a decrease of
$h_{\omega_{N^2},{\cal W}_\infty}(1,\Lambda,n)$;
\item for $n\in \pq{6,10}$ the light--grey region
becomes so elongated that it starts feeling the folding condition so that,
with increasing time--steps, it eventually fully covers the originally 
dark space. In this case, the behavior of
$h_{\omega_{N^2},{\cal W}_\infty}(1,\Lambda,n)$ remains the same
as before up to $n=10$;
\item for $n=11$, $\Gamma_{\Lambda,\alpha}^{(n),N}$
coincides with the whole lattice;
\item for larger times, the frequencies  $\nu_{\Lambda,1}^{(n),N}$ tend
to the constant value $\frac{1}{N^2}$ on almost every point of the
grid. In this case, the behaviour of the entropy production undergoes
a critical change (the crossover occurring at $n=11$) as
showed in fig.~\ref{cinque}.
\end{itemize}
Again, we cannot conclude that $n=11$ is a realistic breaking--time,
because once more we have strong dependence on the chosen partition
(namely from the number $D$ of its elements). For instance, in
fig.~\ref{cinque}, one can see that partitions with $3$ points 
reach their corresponding ``breaking--times''
faster than that with $D=5$; also they do it in an $N$--dependent way.

\begin{figure}[h]
\begin{center}
\ \\[1.5ex]
\begin{picture}(130,70)(0,0) 
\put(10,80){\epsfig{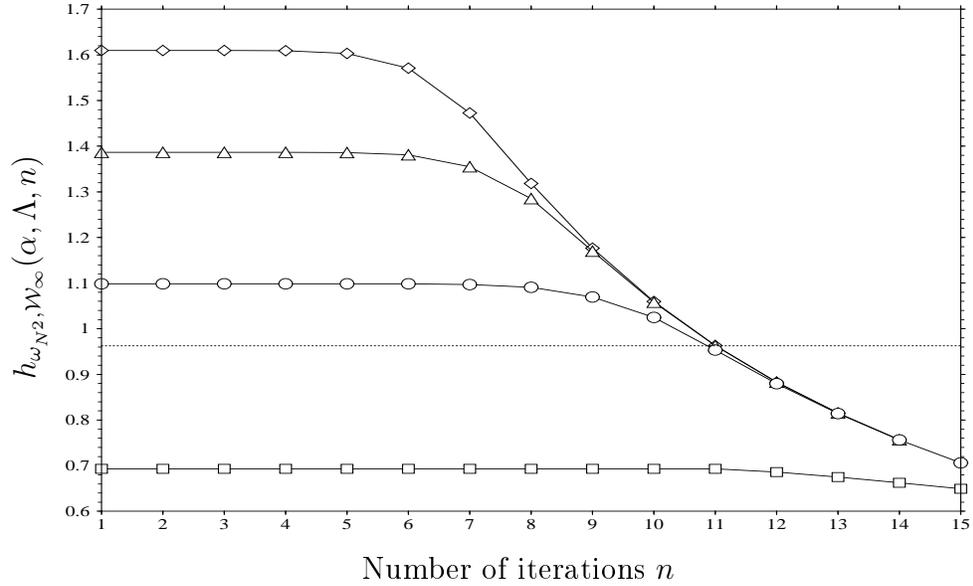}}
\begin{sideways}
\put(10,0){\makebox(70,10)[cc]{$\displaystyle h_{\omega_{N^2},{\cal W}_\infty}(\alpha,\Lambda,n)$}}
\end{sideways}
\put(0,0){\makebox(120,10)[cc]{Number of iterations $n$}}
\end{picture}
\caption{ Entropy production $\displaystyle h_{\omega_{N^2},{\cal
W}_\infty}(\alpha,\Lambda,n)$ 
in four hyperbolic ($\alpha = 1$) cases, for $D$
randomly distributed $\bs{r}_i$ in 
$\Lambda$, with $N=200$. Values for $D$ are: $\diamond = 5$,
\mbox{{\scriptsize 
$\bigtriangleup$} $= 4$}, $\circ = 3$ and {\scriptsize $\Box$} $=
2$. The dotted line corresponds to the Lyapounov exponent $\log\lambda =
0.962\ldots$ at $\alpha = 1$.}
\ 
\label{tre}
\end{center}
\end{figure}

For a chosen set $\Lambda$ consisting of $D$ elements
very close to 
each other 
and $N$ very large,
$h_{\omega_{N^2},{\cal W}_\infty}(\alpha,\Lambda,n)\simeq\log\lambda$
(which is the asymptote in the continuous case)
from a certain $\bar{n}$ up to a time $\tau_B$. Since this latter is
now partition independent, it can
properly be
considered as the
\underline{breaking--time} of the system; it is 
given by
\begin{equation}
\tau_B={\log}_{\lambda}N^2 \label{bbbttt}\cdot
\end{equation}
\begin{center}
\framebox[1.1\width][c]{
\begin{minipage}[c]{10cm}
\enfasi{\ \\[1ex]
It is evident from equation~\eqref{bbbttt} that if one knows
$\tau_B$ then also $\log\lambda$ is known. Usually, one is interested
in the latter which is a sign of the instability of the continuous
classical system. In the following we develop an algorithm which
allows us to extract $\log\lambda$ from studying the corresponding
discretized classical system and its ALF--entropy.\\[0.8ex]
\ }
\end{minipage}}
\end{center}
In working conditions, $N$ is not large enough to allow for $\bar{n}$
being smaller than $\tau_B$; what happens in such a case
is that $H_{\omega_{N^2}}\pt{\alpha,\Lambda,n}\simeq 2\log N$ before
the asymptote for $h_{\omega_{N^2},{\cal
W}_\infty}(\alpha,\Lambda,n)$ is reached. Given $h_{\omega_{N^2},{\cal
W}_\infty}(\alpha,\Lambda,n)$ for $n<\tau_B$, it is thus necessary to seek
means how to estimate the long time behaviour that one would have if
the system were continuous.
\begin{quote}
\begin{NNS}{}\ \label{rem_51}\\[-0.5ex]
When estimating Lyapounov exponents from
discretized hyperbolic classical systems, by using
partitions consisting of nearest neighbors, we have to take into
account some facts:
\begin{Ventry}{\mdseries a.}
\item[\mdseries a.] $h_{\omega_{N^2},{\cal
W}_\infty}(\alpha,\Lambda,n)$ does not increase with $n$; therefore, if
$D<\lambda$, $h_{\omega_{N^2},{\cal
W}_\infty}$ cannot reach the
Lyapounov exponent. 
Denoted by \framebox{$\log\lambda\pt{D}$} the asymptote that we
extrapolate from the data~\footnote{$h_{\omega_{N^2},{\cal
W}_\infty}(\alpha,\Lambda,n)$ may even equal $\log\lambda\pt{D}$ from
the start.}, in general we have \mbox{$\lambda\pt{D}\leqslant\log
D<\lambda$.} 
For instance, for
$\alpha=1$, $\lambda=2.618\ldots>2$ and partitions with \mbox{$D=2$} cannot
produce an entropy greater then $\log 2$; this is the case for the entropies
below the dotted line in figs.~\ref{due} e~\ref{tre}; 
\item[\mdseries b.] partitions with $D$ small but greater than
$\lambda$ allow $\log\lambda$ to be reached in a very short time and 
$\lambda\pt{D}$ is very close to $\lambda$ in this case;
\item[\mdseries c.] partitions with $D\gg\lambda$ require very
long time to converge to $\log\lambda$ (and so very large $N$) and,
moreover, it is not a trivial task to deal with them from a
computational point of view. On the contrary the entropy behaviour for
such partitions offers very good estimates of $\lambda$
(compare, in 
fig.~\ref{cinque}, $\triangleright$ with $\diamond$, {\scriptsize
$\bigtriangleup$}, $\circ$ and {\scriptsize $\Box$}) ;
\item[\mdseries d.] in order to compute $\lambda$ (and then $\tau_B$,
by~\eqref{bbbttt}), one can calculate $\lambda\pt{D}$ for increasing
$D$, until it converges to a stable value $\lambda$;
\item[\mdseries e.] due to number theoretical reasons, the UMG on
${\pt{\IZ/N\IZ}}^2$ present several anomalies. An instance of them 
is showed in fig.~\ref{vicini}
(col. f), where a partition with five nearest neighbors on a lattice
of $200\times200$ points confines the image of
$\bs{f}_{\Lambda,\alpha}^{(n),N}$ (under the 
action of a $T_{\alpha}$ map with $\alpha=17$) on a subgrid of the
torus. In this and analogous cases, there occurs an anomalous depletion of
the entropy production and no significant information is obtainable
from it.
To avoid this difficulties, in
Section~\ref{Sawtooth}
we will go beyond the UMG subclass considered so far. 
\end{Ventry} 
\end{NNS}
\end{quote}
\noindent
\subsubsection{Elliptic regime ($\alpha\in\pg{-1,-2,-3}$)}
One can show that 
all evolution matrices $T_{\alpha}$ are
characterized by the following property:
\begin{equation}
T_{\alpha}^2 = \bar{\alpha} \; T_{\alpha} - \Id \qquad,\qquad
\bar{\alpha}\coleq\pt{\alpha+2} \ \cdot\label{ell_1}
\end{equation}
In the elliptic regime $\alpha\in\pg{-1,-2,-3}$, therefore
$\bar{\alpha}\in\pg{-1, 0, 1}$ and 
relation~\eqref{ell_1} determines a periodic 
evolution with periods:
\begin{subequations}
\label{ell_2}
\begin{alignat}{2}
T_{-1}^{\phantom{-}3} & = -\Id & \qquad\qquad ( T_{-1}^{\phantom{-}6}
& = \Id )\label{ell_2a}\\ 
T_{-2}^{\phantom{-}2} & = -\Id & ( T_{-2}^{\phantom{-}4} & = \Id
)\label{ell_2b}\\  
T_{-3}^{\phantom{-}3} & = +\Id \label{ell_2c}\ \cdot&&
\end{alignat}
\end{subequations}
It has to be stressed that, in the elliptic regime, 
the relations~\eqref{ell_2} \underline{do not hold}
``modulo $N$'', instead they are completely independent from $N$.

Due to the high degree of symmetry in
relations (\ref{ell_1}--\ref{ell_2}), the frequencies
$\nu_{\Lambda,\alpha}^{(n),N}$ 
are different from zero only on a small subset of the whole lattice.

\begin{figure}[h]
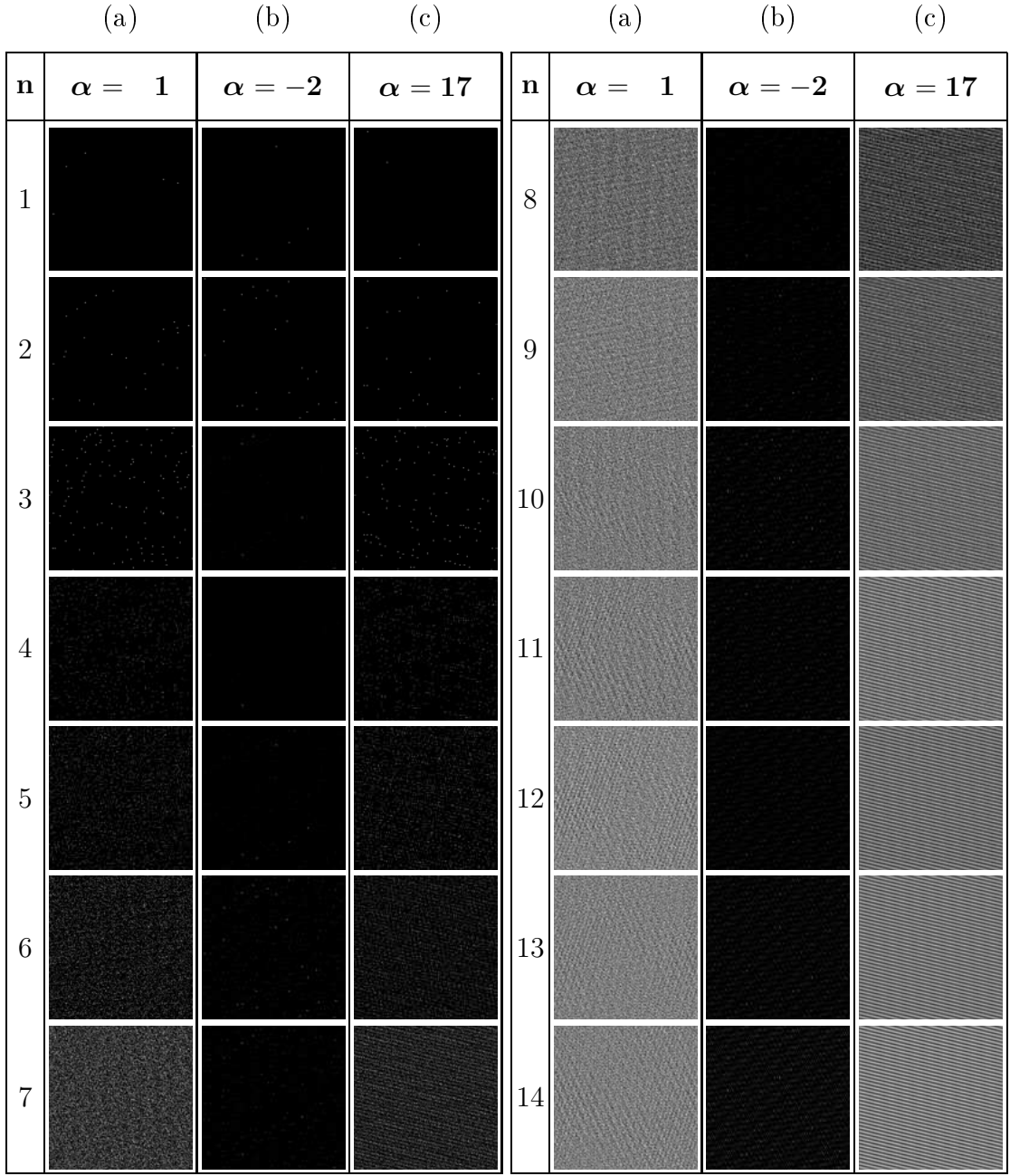

\begin{center}
\begin{picture}(140,164)(0,0) 
\put(0,132){\epsfig{width=\fotogram,file=a_01.epsi}
\epsfig{width=\fotogram,file=b_01.epsi}
\epsfig{width=\fotogram,file=c_01.epsi}\hspace{\sepfoto}
\epsfig{width=\fotogram,file=a_08.epsi}
\epsfig{width=\fotogram,file=b_08.epsi}
\epsfig{width=\fotogram,file=c_08.epsi}}
\put(0,110){\epsfig{width=\fotogram,file=a_02.epsi}
\epsfig{width=\fotogram,file=b_02.epsi}
\epsfig{width=\fotogram,file=c_02.epsi}\hspace{\sepfoto}
\epsfig{width=\fotogram,file=a_09.epsi}
\epsfig{width=\fotogram,file=b_09.epsi}
\epsfig{width=\fotogram,file=c_09.epsi}}
\put(0,88){\epsfig{width=\fotogram,file=a_03.epsi}
\epsfig{width=\fotogram,file=b_03.epsi}
\epsfig{width=\fotogram,file=c_03.epsi}\hspace{\sepfoto}
\epsfig{width=\fotogram,file=a_10.epsi}
\epsfig{width=\fotogram,file=b_10.epsi}
\epsfig{width=\fotogram,file=c_10.epsi}}
\put(0,66){\epsfig{width=\fotogram,file=a_04.epsi}
\epsfig{width=\fotogram,file=b_04.epsi}
\epsfig{width=\fotogram,file=c_04.epsi}\hspace{\sepfoto}
\epsfig{width=\fotogram,file=a_11.epsi}
\epsfig{width=\fotogram,file=b_11.epsi}
\epsfig{width=\fotogram,file=c_11.epsi}}
\put(0,44){\epsfig{width=\fotogram,file=a_05.epsi}
\epsfig{width=\fotogram,file=b_05.epsi}
\epsfig{width=\fotogram,file=c_05.epsi}\hspace{\sepfoto}
\epsfig{width=\fotogram,file=a_12.epsi}
\epsfig{width=\fotogram,file=b_12.epsi}
\epsfig{width=\fotogram,file=c_12.epsi}}
\put(0,22){\epsfig{width=\fotogram,file=a_06.epsi}
\epsfig{width=\fotogram,file=b_06.epsi}
\epsfig{width=\fotogram,file=c_06.epsi}\hspace{\sepfoto}
\epsfig{width=\fotogram,file=a_13.epsi}
\epsfig{width=\fotogram,file=b_13.epsi}
\epsfig{width=\fotogram,file=c_13.epsi}}
\put(0,0){\epsfig{width=\fotogram,file=a_07.epsi}
\epsfig{width=\fotogram,file=b_07.epsi}
\epsfig{width=\fotogram,file=c_07.epsi}\hspace{\sepfoto}
\epsfig{width=\fotogram,file=a_14.epsi}
\epsfig{width=\fotogram,file=b_14.epsi}
\epsfig{width=\fotogram,file=c_14.epsi}}

\put(-6.42,164){\line(0,-6){164.7}}
\put(-0.82,164){\line(0,-6){164.7}}
\put(21.58,164){\line(0,-6){164.7}}
\put(43.98,164){\line(0,-6){164.7}}
\put(66.38,164){\line(0,-6){164.7}}
\put(67.78,164){\line(0,-6){164.7}}
\put(73.38,164){\line(0,-6){164.7}}
\put(95.78,164){\line(0,-6){164.7}}
\put(118.18,164){\line(0,-6){164.7}}
\put(140.58,164){\line(0,-6){164.7}}
\put(-6.48,164){\line(6,0){72.94}}
\put(-6.48,154){\line(6,0){72.94}}
\put(-6.48,-0.7){\line(6,0){72.94}}
\put(67.72,164){\line(6,0){72.94}}
\put(67.72,154){\line(6,0){72.94}}
\put(67.72,-0.7){\line(6,0){72.94}}
\put(-6.42,154){\makebox(5.6,10)[cc]{$\mathbf{n}$}}
\put(-0.82,154){\makebox(22.4,10)[cc]{$\boldsymbol{\alpha}\ \mathbf{= \phantom{-}1}$}}
\put(21.58,154){\makebox(22.4,10)[cc]{$\boldsymbol{\alpha}\ \mathbf{= -2}$}}
\put(43.98,154){\makebox(22.4,10)[cc]{$\boldsymbol{\alpha}\ \mathbf{= 17}$}}
\put(-0.82,164){\makebox(22.4,10)[cc]{(a)}}
\put(21.58,164){\makebox(22.4,10)[cc]{(b)}}
\put(43.98,164){\makebox(22.4,10)[cc]{(c)}}
\put(67.78,154){\makebox(5.6,10)[cc]{$\mathbf{n}$}}
\put(73.38,154){\makebox(22.4,10)[cc]{$\boldsymbol{\alpha}\ \mathbf{= \phantom{-}1}$}}
\put(95.78,154){\makebox(22.4,10)[cc]{$\boldsymbol{\alpha}\ \mathbf{= -2}$}}
\put(118.18,154){\makebox(22.4,10)[cc]{$\boldsymbol{\alpha}\ \mathbf{= 17}$}}
\put(73.38,164){\makebox(22.4,10)[cc]{(a)}}
\put(95.78,164){\makebox(22.4,10)[cc]{(b)}}
\put(118.18,164){\makebox(22.4,10)[cc]{(c)}}
\put(-6.42,131.5){\makebox(5.6,22)[cc]{$1$}}
\put(-6.42,109.5){\makebox(5.6,22)[cc]{$2$}}
\put(-6.42,87.5){\makebox(5.6,22)[cc]{$3$}}
\put(-6.42,65.5){\makebox(5.6,22)[cc]{$4$}}
\put(-6.42,43.5){\makebox(5.6,22)[cc]{$5$}}
\put(-6.42,21.5){\makebox(5.6,22)[cc]{$6$}}
\put(-6.42,-0.5){\makebox(5.6,22)[cc]{$7$}}
\put(67.78,131.5){\makebox(5.6,22)[cc]{$8$}}
\put(67.78,109.5){\makebox(5.6,22)[cc]{$9$}}
\put(67.78,87.5){\makebox(5.6,22)[cc]{$10$}}
\put(67.78,65.5){\makebox(5.6,22)[cc]{$11$}}
\put(67.78,43.5){\makebox(5.6,22)[cc]{$12$}}
\put(67.78,21.5){\makebox(5.6,22)[cc]{$13$}}
\put(67.78,-0.5){\makebox(5.6,22)[cc]{$14$}}
\end{picture}
\caption{Density plots showing the frequencies 
$\displaystyle \nu_{\Lambda,\alpha}^{(n),N}$
in two hyperbolic regimes (columns a and c) and
an elliptic one (col. b),
for five randomly distributed $\bs{r}_i$ in
$\Lambda$ with $N=200$. Black corresponds to
$\displaystyle \nu_{\Lambda,\alpha}^{(n),N} = 0$.
In the hyperbolic cases, $\displaystyle \nu_{\Lambda,\alpha}^{(n),N}$
tends to
equidistribute on ${\pt{\IZ/N\IZ}}^2$ with
increasing $n$ and becomes constant 
when the breaking--time is reached.}
\label{lontani}
\end{center}
\end{figure}

\begin{figure}[h]
\begin{center}
\ \\[2ex]
\begin{picture}(130,70)(0,0) 
\put(10,80){\epsfig{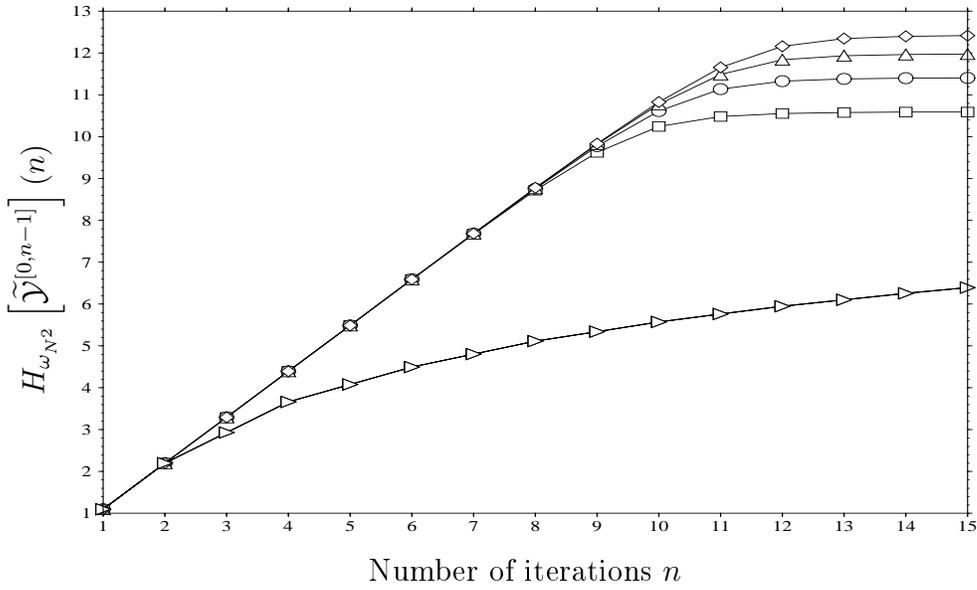}}
\begin{sideways}
\put(10,0){\makebox(70,10)[cc]{$\displaystyle H_{\omega_{N^2}}\pq{{\cal \widetilde
Y}^{[0,n-1]}}\pt{n}$}}
\end{sideways}
\put(0,0){\makebox(120,10)[cc]{Number of iterations $n$}}
\end{picture}
\caption{Von Neumann entropy $\displaystyle H_{\omega_{N^2}}\pt{n}$ 
in four hyperbolic ($\alpha = 1$ for $\diamond$, {\scriptsize
$\bigtriangleup$}, $\circ$, {\scriptsize $\Box$}) and
four elliptic ($\alpha = -2$ for $\triangleright$) cases, for three
randomly distributed $\bs{r}_i$ in 
$\Lambda$. Values for $N$ are: $\diamond = 500$, {\scriptsize
$\bigtriangleup$} $= 400$, $\circ = 300$ and {\scriptsize $\Box$} $=
200$, whereas the curve labeled by $\triangleright$ represents
four elliptic systems with $N\in\pg{200, 300, 400, 500}$.}
\ 
\label{uno}
\end{center}
\end{figure}

This behavior is apparent in fig.~\ref{lontani}~:~col. b, in which
we consider five randomly distributed $\bs{r}_i$ in $\Lambda$, and in
fig.~\ref{vicini}~:~col. e, in which the five $\bs{r}_i$ are grouped as
in fig.~\ref{lattice}. In both cases, the Von Neumann entropy
$\displaystyle H_{\omega_{N^2}}\pt{n}$ is not
linearly increasing with $n$ (see figure~\ref{quattro}), instead it assumes a
$\log$--shaped
profile ( up to the breaking--time, see fig.~\ref{uno}).\\[-2ex]
\begin{quote}
\begin{NNN}{}\ \\[-0.5ex]
The last observation indicates how the entropy production analysis can be
used to recognize whether a dynamical systems is hyperbolic or not.
If we use randomly
distributed points as a partition, we observe that hyperbolic systems
show constant entropy production (up to the breaking--time),
whereas the others do not.

Moreover, unlike hyperbolic ones, elliptic systems do not change their
behaviour with $N$ (for 
reasonably large $N$) as clearly showed in fig.~\ref{uno}, in which
elliptic systems ($\alpha=-2$) with four different values of $N$ give
the same plot.
On the contrary, we have dependence on how rich is the chosen
partition, similarly to 
what we have for hyperbolic systems, as showed in
fig.~\ref{cinque}.
\end{NNN}
\end{quote}
\noindent
\subsubsection{Parabolic regime ($\alpha\in\pg{0, 4}$)}
This regime is characterized by  $\lambda = \lambda^{-1} = \pm 1$,
that is $\log\abs{\lambda} = 0$ (see Remark~\ref{Rem_21}, c.).
These systems behave as the hyperbolic ones
(see subsections~\ref{sez_511} and ~\ref{sub_512}) and this is true also
for the the general behavior of the entropy production, apart from the
fact that we never fall in the condition (a.) of Remark~\ref{rem_51}.
Then, for sufficiently large $N$, every partition consisting of $D$
grouped $\bs{r}_i$ will reach the asymptote $\log\abs{\lambda}=0$.
\subsection{The case of Sawtooth Maps} \label{Sawtooth}
The Sawtooth Maps~\cite{Ref_24,Ref_25},
are triples $({\cal X},\mu,S_\alpha)$ where
\begin{subequations}
\label{SSAoDC_1}
\begin{align}
{\cal X}&={\IT}^2={\IR}^2/{\IZ}^2=\left\{\bs{x}=(x_1,x_2)\ \pmod{1} \right\}
\label{SSAoDC_1a}\\
S_\alpha 
\begin{pmatrix}
x_1\\
x_2
\end{pmatrix}& =
\begin{pmatrix}
1+\alpha & 1\\
\alpha & 1
\end{pmatrix}
\begin{pmatrix}
\bk{x_1}\\
x_2
\end{pmatrix}
\ \pmod{1}\ ,\quad
\alpha\in\IR
\label{SSAoDC_1b}\\
\ud\mu(\bs{x})&=\ud x_1\; \ud x_2 \ ,
\label{SSAoDC_1c}
\end{align} 
\end{subequations}
where $\bk{\cdot}$ denotes
the fractional part of a real number. Without
$\bk{\cdot}$,~\eqref{SSAoDC_1b} is not well defined on ${\IT}^2$ for
not--integer $\alpha$; in fact, without taking the fractional part,
the same point $\bs{x} = \bs{x} + \bs{n} \in {\IT}^2 , \bs{n} \in
{\IZ}^2$, would have (in general) $S_\alpha \pt{\bs{x}} \neq S_\alpha
\pt{\bs{x}+\bs{n}}$. Of course, $\bk{\cdot}$ is not necessary when
$\alpha\in{\IZ}$.\\
The Lebesgue measure defined in~\eqref{SSAoDC_1c} is \enfasi{invariant}
for all $\alpha\in\IR$.\\
After identifying $\bs{x}$ with canonical coordinates
$(q,p)$ and imposing the$\pmod{1}$ condition on
both of them, the above dynamics can be rewritten as:
\begin{equation}
\begin{cases}
q^\prime &= q + p^\prime\\
p^\prime &= p + \alpha  \bk{q}
\end{cases}
\pmod{1},
\label{SSAoDC_11}
\end{equation}
This is nothing but the Chirikov Standard Map~\cite{Ref_4} in which
$-\frac{1}{2\pi}\sin(2\pi q)$ is replaced by 
$\bk{q}$.
The dynamics in~\eqref{SSAoDC_11} can also be thought of as generated
by the (singular) Hamiltonian 
\begin{equation}
H(q,p,t)=\frac{p^2}{2}-\alpha\, 
\frac{{\bk{q}}^2}{2}\,\delta_p(t),
\end{equation}
where $\delta_p(t)$ is the periodic Dirac delta which makes the
potential act through
periodic kicks with period 
$1$.

\begin{figure}[h]
\begin{center}
\begin{picture}(130,70)(0,0) 
\put(10,80){\epsfig{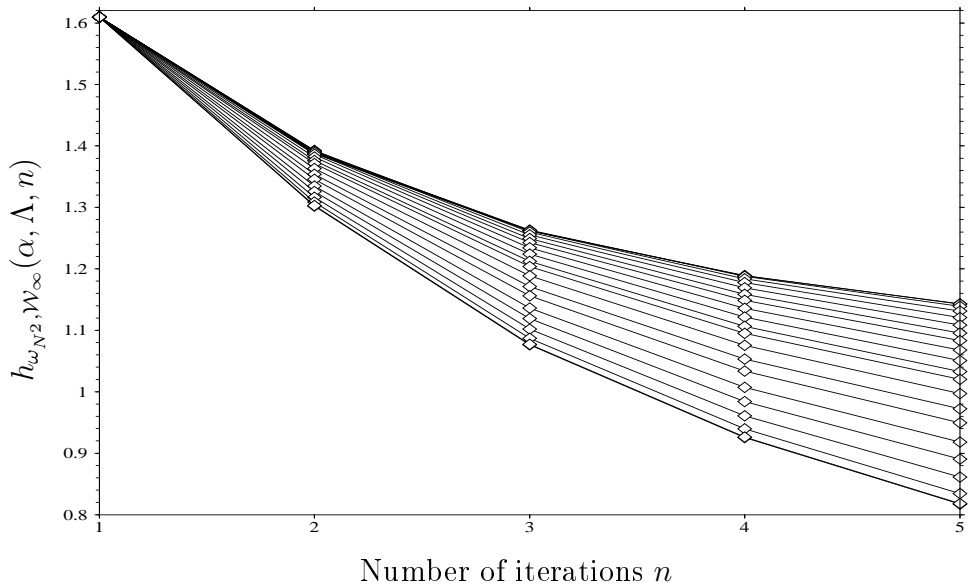}}
\begin{sideways}
\put(10,0){\makebox(70,10)[cc]{$\displaystyle h_{\omega_{N^2},{\cal
W}_\infty}(\alpha,\Lambda,n)$}} 
\end{sideways}
\put(0,0){\makebox(120,10)[cc]{Number of iterations $n$}}
\end{picture}
\caption{Entropy production $\displaystyle h_{\omega_{N^2},{\cal
W}_\infty}(\alpha,\Lambda,n)$  for $21$
hyperbolic Sawtooth maps, relative to a 
for a cluster of $5$ nearest neighborings points $\bs{r}_i$ in 
$\Lambda$, with $N=38$.
The parameter $\alpha$ decreases from $\alpha=1.00$
(corresponding to 
the upper curve) to $\alpha=0.00$ (lower curve) through
$21$ equispaced steps.}
\label{sei}
\end{center}
\end{figure}

\begin{figure}[h]
\begin{center}
\begin{picture}(130,70)(0,0) 
\put(10,80){\epsfig{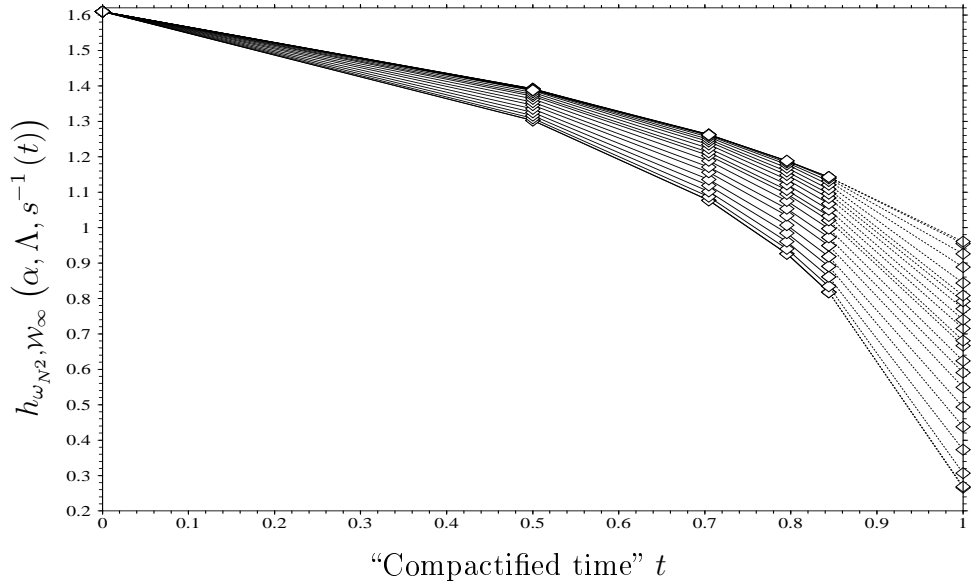}}
\begin{sideways}
\put(10,0){\makebox(70,10)[cc]{$\displaystyle h_{\omega_{N^2},{\cal
W}_\infty}\pt{\alpha,\Lambda,s^{-1}\pt{t}}$}}
\end{sideways}
\put(0,0){\makebox(120,10)[cc]{``Compactified time'' $t$}}
\end{picture}
\caption{The solid lines correspond to $\displaystyle
{\pt{s_n\,,\,h_{\omega_{N^2},{\cal 
W}_\infty}(\alpha,\Lambda,n)}}$, with $n\in\pg{1,2,3,4,5}$, for the
values of $\alpha$ considered in figure~\ref{sei}. Every
$\alpha$--curve is continued as a dotted line
up to $\pt{1,l^5_\alpha}$, where
$l^5_\alpha$ is the Lyapounov exponent extracted from the curve by
fitting all the five points via a Lagrange polynomial ${\cal P}^m\pt{t}$.}
\
\label{sette}
\end{center}
\end{figure}

\begin{figure}[h]
\begin{center}
\begin{picture}(130,70)(0,0) 
\put(10,80){\epsfig{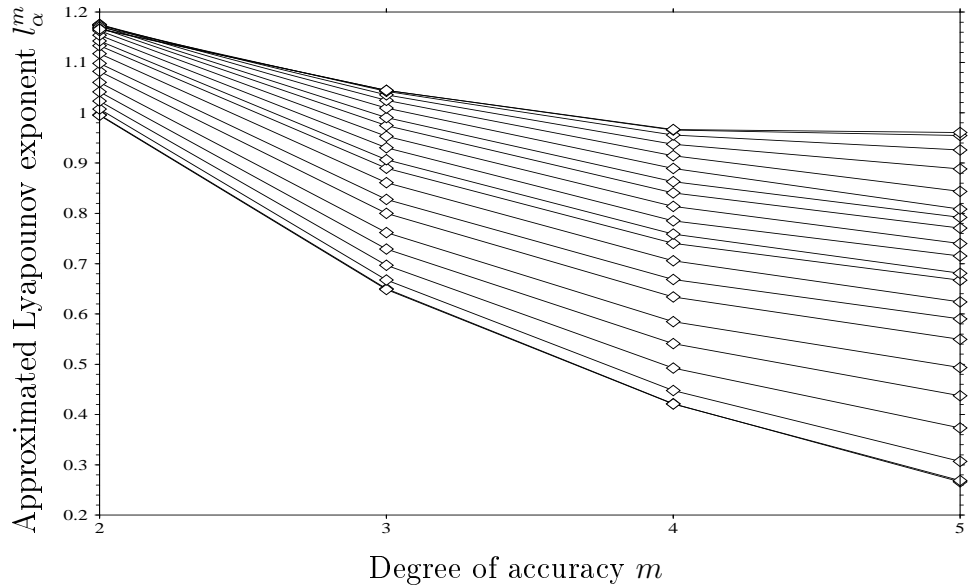}}
\begin{sideways}
\put(10,0){\makebox(70,10)[cc]{Approximated Lyapounov exponent $l^m_\alpha$}}
\end{sideways}
\put(0,0){\makebox(120,10)[cc]{Degree of
accuracy $m$}}
\end{picture}
\caption{Four estimated Lyapounov exponents $l^m_\alpha$ 
plotted vs. their degree of accuracy $m$ for the values of
$\alpha$ considered in figures~\ref{sei}~and~\ref{sette}.}
\label{otto}
\end{center}
\end{figure}

\begin{figure}[h]
\begin{center}
\begin{picture}(130,70)(0,0) 
\put(10,80){\epsfig{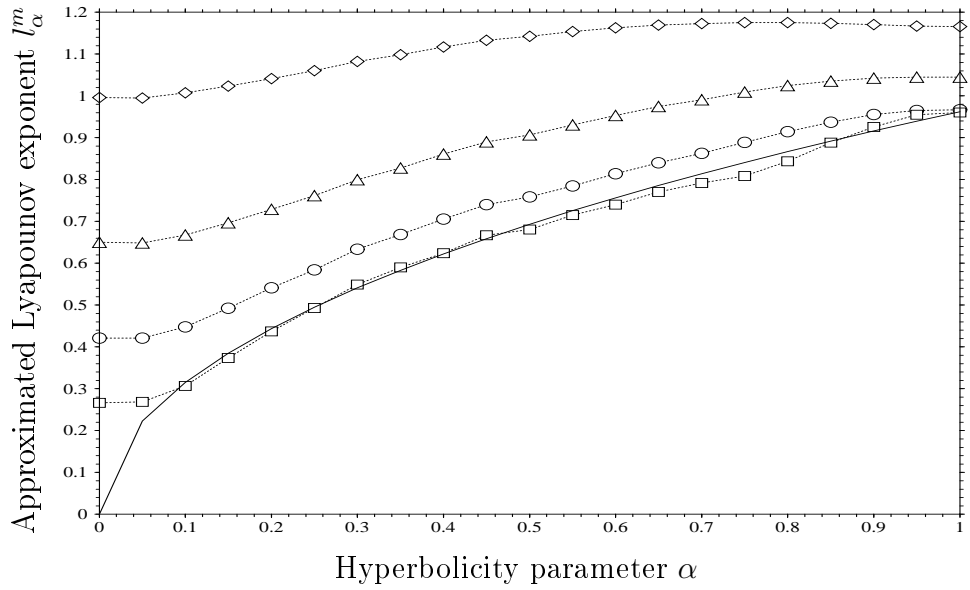}}
\begin{sideways}
\put(10,0){\makebox(70,10)[cc]{Approximated Lyapounov exponent
$l^m_\alpha$}}
\end{sideways}
\put(0,0){\makebox(120,10)[cc]{Hyperbolicity parameter $\alpha$}}
\end{picture}
\caption{Plots of the four estimated of Lyapounov exponents $l^m_\alpha$ of
figure~\ref{otto} 
vs. the considered values of
$\alpha$.
The polynomial degree $m$ is as follows: $\diamond = 2$,
\mbox{{\scriptsize 
$\bigtriangleup$} $= 3$}, $\circ = 4$ and {\scriptsize $\Box$} $=
5$. The solid line corresponds to the theoretical Lyapounov exponent 
$\displaystyle \log\lambda_\alpha=
\log\:(\alpha+2 + \sqrt{\alpha\:(\alpha+4)\,}\;) - \log 2
$.}
\label{nove}
\end{center}
\end{figure}

\noindent Sawtooth Maps are invertible and the inverse is given by the
expression   
\begin{gather}
S_\alpha^{-1}
\begin{pmatrix}
x_1\\
x_2
\end{pmatrix} =
\begin{pmatrix}
\phantom{-}1 & 0\\
-\alpha & 1
\end{pmatrix}
\bk{\begin{pmatrix}
1 & -1\\
0 & \phantom{-}1
\end{pmatrix}
\begin{pmatrix}
x_1\\
x_2
\end{pmatrix}}
\ \pmod{1}
\label{SSAoDC_1d}\\
\intertext{or, in other words,}
\begin{cases}
q &= \phantom{-\alpha\,}q^\prime - p^\prime\\
p &= -\alpha\,q^{\phantom{\prime}} + p^\prime
\end{cases}
\pmod{1}\ .
\label{SSAoDC_1e}
\end{gather}
It can indeed be checked that
$S_\alpha\pt{S_\alpha^{-1}\pt{\bs{x}}} =
S_\alpha^{-1}\pt{S_\alpha\pt{\bs{x}}}=\bk{\bs{x}},\ \forall
\bs{x}\in{\IT}^2$.\\[-2ex]
\begin{quote}
\begin{NNS}{}\ \\[-5ex]\label{SSRem_21}
\begin{Ventry}{\mdseries viii.}
 \item[\mdseries i.] Sawtooth Maps $\{S_\alpha\}$ are
	\enfasi{discontinuous} on
	the subset \\$\gamma_0\coleq\pg{\bs{x} = \pt{0,p},\
	p\in{\IT}}\in {\IT}^2$: 
	two points close to this border,
	\mbox{$A\coleq\pt{\varepsilon,p}$} and
	$B\coleq\pt{1-\varepsilon,p}$,
	have images that differ, in the $\varepsilon \rightarrow 0$ limit, by a
	vector $d^{(1)}_{S_{\alpha}^{\phantom{-1}}}(A,B)=\pt{\alpha,\alpha} \pmod{1}$.
\item[\mdseries ii.] Inverse Sawtooth Maps $\{S_\alpha^{-1}\}$ are
	\enfasi{discontinuous} on
	the subset \\$\gamma_{-1}\coleq S_\alpha\pt{\gamma_0} =
	\pg{\bs{x} = \pt{p,p},\
	p\in{\IT}}\in {\IT}^2$:
	two points close to this border,
	\mbox{$A\coleq\pt{p+\varepsilon,p-\varepsilon}$} and
	$B\coleq\pt{p-\varepsilon,p+\varepsilon}$,
	have images that differ, in the $\varepsilon \rightarrow 0$ limit, by a
	vector $d^{(1)}_{S_{\alpha}^{-1}}(A,B)=\pt{0,\alpha} \pmod{1}$.
\item[\mdseries iii.] The hyperbolic, elliptic or parabolic behavior
of Sawtooth 
        maps is related to the eigenvalues of 
	$\pt{\begin{smallmatrix} 1+ \alpha  & 1\\ \alpha & 1 
	\end{smallmatrix}}$ 
 exactly as in 
 	Remark~\ref{Rem_21}.ii.
\item[\mdseries iv.] The Lebesgue measure in~\eqref{SSAoDC_1c} is
	$S_\alpha^{-1}$--invariant. 
\end{Ventry}
\end{NNS}
\end{quote}
\noindent
From a computational point of view,
the study of the entropy production in the case of Sawtooth Maps
${S_{\alpha}}$ is more
complicated than for the
${T_{\alpha}}$'s.
The reason to study numerically these dynamical systems is twofold:
\begin{itemize}
	\item to avoid the difficulties described in Remark~\ref{rem_51}~(e.);
	\item to deal, in a way compatible with numerical computation
	limits, with the largest possible spectrum of accessible
	Lyapounov exponent. We know that for
	$\alpha\in\IZ\bigcap\pg{\text{non elliptic domain}}$,
	\begin{equation*}
	\displaystyle \lambda^{\pm}\pt{T_{\alpha}} =
	\lambda^{\pm}\pt{S_{\alpha}} = 
	\frac{\alpha+2\pm\sqrt{(\alpha+2)^2-4}}{2}\ \cdot
	\end{equation*}
	In order to fit $\log\lambda_\alpha$ (
	$\log\lambda_\alpha$ being 
	the Lyapounov exponent corresponding to a
	given $\alpha$) via entropy production analysis, we need $D$
	elements in the partition (see points b. and c. of
	Remark~\ref{rem_51}) with $D\geq\lambda_\alpha$.
	Moreover, if we were to
	study the power of our method for \underline{different
	integer} values of $\alpha$ we would be forced 
	forced to use very large $D$, in which case we would need
	very long computing times in order to evaluate numerically the
	entropy production $h_{\omega_{N^2},{\cal
	W}_\infty}(\alpha,\Lambda,n)$ in a reasonable interval of times
	$n$. 
	Instead, for Sawtooth Maps, we can fix the parameters
	$\pt{N,D,\Lambda}$ and study $\lambda_\alpha$ for $\alpha$
	confined in a small domain, but free to assume every real
	value in that domain.
\end{itemize}
In the following, we investigate the case of $\alpha$ in the hyperbolic
regime  
with $D$ nearest neighbors $\bs{r}_i$ in $\Lambda$, as done in
subsection~\ref{sub_512}. In particular, figures
(\ref{sei}--\ref{nove}) refer to the following fixed parameters:\\[-6ex]
\begin{center}
\begin{tabular}{%
@{}c
@{}c
@{}p{6mm}
@{}c
@{}c
@{}c
@{}p{6mm}
@{}c
@{}c
@{}c
@{}p{6mm}
@{}c
@{}c
@{}c
@{}p{6mm}
@{}c
@{}c
@{}c
@{}p{6mm}
@{}c
@{}c
@{}c
@{}}
\multicolumn{2}{c}{\rule[-1ex]{0pt}{6ex}} &
$N$ & $=$ & $38$ & 
\multicolumn{2}{c}{\quad} &
$;$ & \ &
\multicolumn{2}{c}{$n_{\text{max}}$} 
& $=$ & $5$ &
\multicolumn{2}{c}{\quad} &
$;$ & 
\multicolumn{2}{c}{\quad} &
$D$ & $=$ & $5$ &
$\;\;;$\\
$\Lambda$\rule[-1ex]{0pt}{7ex} & $\quad : \quad$ &
$\bs{r}_1$ & $=$ & $\displaystyle\begin{pmatrix}7\\8\end{pmatrix}$ & $\;\;,\;\;$ &
$\bs{r}_2$ & $=$ & $\displaystyle\begin{pmatrix}7\\9\end{pmatrix}$ & $\;\;,\;\;$ &
$\bs{r}_3$ & $=$ & $\displaystyle\begin{pmatrix}6\\8\end{pmatrix}$ & $\;\;,\;\;$ &
$\bs{r}_4$ & $=$ & $\displaystyle\begin{pmatrix}7\\7\end{pmatrix}$ & $\;\;,\;\;$ &
$\bs{r}_5$ & $=$ & $\displaystyle\begin{pmatrix}8\\8\end{pmatrix}$ & $\;\;;$\\
$\alpha$ \rule[-1ex]{0pt}{6ex}& $\quad : \quad$ &
\multicolumn{19}{c}{from $0.00$ to $1.00$ with an incremental step
	of $0.05$.} &
\end{tabular}\\[4ex]
\end{center}
First, we compute the Von Neumann entropy~\eqref{CoAFE_10}
using the (hermitian) matrix ${\cal
G}_{\bs{\ell}_1,\bs{\ell}_2}\pt{n}$ defined in 
~\eqref{CoAFE_1803}. This is actually a diagonalization problem:
once that the $N^2$ eigenvalues ${\pg{\eta_i}}_{i=1}^{N^2}$
are found, then 
\begin{equation}
H_{\omega_{N^2}}\pq{{\cal \widetilde Y}^{[0,n-1]}} = - \sum_{i=1}^{N^2}
\eta_{i}\log \eta_{i} \ \cdot\label{swtm_1} 
\end{equation}
Then, from~\eqref{AEP_1}, we can determine $h_{\omega_{N^2},{\cal
W}_\infty}(\alpha,\Lambda,n)$. 
In the numerical example, the ($\Lambda$--dependent) breaking--time
occurs after $n=5$; for this reason we have chosen $n_{\text{max}}=5$.
In fact, we are interested in the region where the discrete
system behaves almost as a continuous one.

In figure~\ref{sei}, the entropy production is plotted for the chosen
set of $\alpha$'s: for very large $N$ (that is close to the
continuum limit, in which no breaking--time occurs) all 
curves (characterized by different $\alpha$'s) 
would tend to \mbox{$\log\lambda_\alpha$ with $n$.}

One way to determine the asymptote $\log\lambda_\alpha$ is to fit the
decreasing function $\displaystyle h_{\omega_{N^2},{\cal
W}_\infty}(\alpha,\Lambda,n)$ over the range of data and
extrapolate the fit for $n\to\infty$. 
Of course we can not perform
the fit with polynomials, because every polynomial diverges in the 
$n\to\infty$ limit.

A better strategy is to compactify the time evolution by means of a
isomorphic positive function $s$ with bounded range, for instance:
\begin{equation}
\IN\ni n  \longmapsto
s_n \coleq \frac{2}{\pi}\arctan\pt{n-1}\in\pq{0,1}\ \cdot
\label{swtm_2} 
\end{equation}
Then, for fixed $\alpha$, in fig.~\ref{sette} we consider 
$n_{\text{max}}$ points
$\displaystyle {\pt{s_n\,,\,h_{\omega_{N^2},{\cal
W}_\infty}(\alpha,\Lambda,n)}}$ and 
extract the 
asymptotic value of $\displaystyle h_{\omega_{N^2},{\cal
W}_\infty}(\alpha,\Lambda,n)$ for $n\to\infty$, that is the value of $\displaystyle h_{\omega_{N^2},{\cal
W}_\infty}\pt{\alpha,\Lambda,s^{-1}\pt{t}}$ for $t\to 1^-$, as
follows.

Given a graph consisting of $m\in\pg{2,3,\cdots,n_{\text{max}}}$
points, in our case
the first $m$ points of 
curves as in fig.~\ref{sette}, namely
\begin{equation*}
\pg{\pt{s_1\,,\,h_{\omega_{N^2},{\cal
W}_\infty}(\alpha,\Lambda,1)},\pt{s_2\,,\,h_{\omega_{N^2},{\cal
W}_\infty}(\alpha,\Lambda,2)}, \cdots,\pt{s_m\,,\,h_{\omega_{N^2},{\cal
W}_\infty}(\alpha,\Lambda,m)}},
\nonumber
\end{equation*}
the data are fit by a Lagrange polynomial ${\cal P}^m \pt{t}$ (of
degree $m-1$) 
\begin{subequations}
\label{swtm_3}
\begin{align}
{\cal P}^m \pt{t} & = \sum_{i=1}^m \;P_i \pt{t}\label{swtm_3a}\\
\text{where}\qquad P_i \pt{t} & = \prod_{\substack{j = 1\\j\neq \,i}}^m
\;\frac{t - s_j}{s_i - s_j} \; h_{\omega_{N^2},{\cal
W}_\infty}(\alpha,\Lambda,i)\ \cdot\label{swtm_3b}
\end{align}
\end{subequations}
The value assumed by this polynomial when $t\to 1^-$ (corresponding to
$n\to\infty$) will be the estimate (of degree $m$) of the
Lyapounov exponent, denoted by $l^m_\alpha$:
the higher the value of $m$, the more 
accurate the estimate. From~\eqref{swtm_3} we get:
\begin{equation}
l^m_\alpha \coleq {\cal P}^m \pt{t} {\Big|}_{t=1} = \sum_{i=1}^m \;
h_{\omega_{N^2},{\cal 
W}_\infty}(\alpha,\Lambda,i)\;
\prod_{\substack{j = 1\\j\neq \,i}}^m 
\;\frac{1 - s_j}{s_i - s_j}\ \cdot
\label{swtm_4}
\end{equation}
The various $l^m_\alpha$ are plotted in figure~\ref{otto}
as functions of $m$ for
all considered $\alpha$. The convergence of
$l^m_\alpha$ with $m$ is showed in figure~\ref{nove},
together with the theoretical Lyapounov exponent $\log\lambda_\alpha$;
as expected, we find that the latter is the asymptote of
${\pg{l^m_\alpha}}_m$ with respect to the polynomial degree $m$. 

The dotted line in fig.~\ref{sette} extrapolates $21\ \alpha$--curves
in compactified time up to $t=1$ using five points
in the Lagrange polynomial approximation.\\[-5ex]
\section{Conclusions}
In this paper, we have considered discretized hyperbolic classical
systems on the torus by forcing them on a squared lattice with spacing
$\frac{1}{N}$. We showed how the discretization procedure is similar to
quantization; in particular, following the analogous case of the
classical limit $\hbar\longmapsto 0$, we have set up the theoretical framework
to discuss the continuous limit $N\longmapsto\infty$. Furthermore,
using the similarities between discretized and quantized classical
systems, we have applied the quantum dynamical entropy of Alicki and
Fannes to study the footprints of classical (continuous) chaos as it
is expected to reveal itself, namely through the presence of
characteristic time scales and corresponding breaking--times. Indeed,
exactly as in quantum chaos, a discretized hyperbolic system can
mimick its continuous partner only up to times which scale as $\log
N$. We have also extended the numerical analysis from the so called
Arnold cat maps to the discontinuous Sawtooth maps, whose
interpretation in the theoretical frame set up in this work will be
discussed in a forthcoming paper.

\noindent
\textbf{\large Acknowledgments}

\noindent This work is supported in part by {\bf CONACyT} project number
U$40004$--F.

\bigskip
\bigskip
%
%
%
%
\newpage
%
%
%
%
%
\newpage
%
%
%
%
\newpage
%
\begin{figure}[h]
\begin{center}
\begin{picture}(130,70)(0,0) 
\put(10,80){\epsfig{width=\altgraph,height=\larggraph,angle=-90,file=quattro.epsi}}
\begin{sideways}
\put(10,0){\makebox(70,10)[cc]{$\displaystyle H_{\omega_{N^2}}\pq{{\cal \widetilde
Y}^{[0,n-1]}}\pt{n}$}}
\end{sideways}
\put(0,0){\makebox(120,10)[cc]{Number of iterations $n$}}
\end{picture}
\caption{ Von Neumann entropy $\displaystyle H_{\omega_{N^2}}\pt{n}$ 
in four elliptic ($\alpha = -2$) cases, for $D$
randomly distributed $\bs{r}_i$ in 
$\Lambda$, with $N=200$. Value for $D$ are: $\diamond = 5$,
\mbox{{\scriptsize 
$\bigtriangleup$} $= 4$}, $\circ = 3$ and {\scriptsize $\Box$} $=
2$.}
\
\label{quattro}
\end{center}
\end{figure}
%
\newpage
%
%
\newpage
%
%

\end{document}